\documentclass[%
 reprint,
 amsmath,amssymb,
 aps,
 prl,
]{revtex4-1}

\usepackage{enumitem}% costume listings
\usepackage{graphicx}% Include figure files
\usepackage{dcolumn}% Align table columns on decimal point
\usepackage{bm}% bold math
\usepackage[unicode=true,pdfusetitle,
 bookmarks=true,bookmarksnumbered=false,bookmarksopen=false,
 breaklinks=true,pdfborder={0 0 0},pdfborderstyle={},backref=false,colorlinks=true]{hyperref}% add hypertext capabilities
\usepackage{lipsum}
\usepackage{xfrac}
\usepackage{mathtools}
\usepackage{braket}
\usepackage{amsmath}
\usepackage{enumitem}
\usepackage{mathdots}
\usepackage{bbm}
\usepackage{adjustbox}

\usepackage{lmodern}
\usepackage[utf8]{inputenc}
\usepackage{float}
\usepackage{units}
\usepackage{amssymb}
\usepackage{cancel}
\usepackage{graphicx}
\usepackage{setspace}
\usepackage{braket}
\usepackage{color}
\usepackage[table]{xcolor}

\usepackage{xcolor,cancel}

\begin{document}

\preprint{APS/123-QED}

\title{Wavefronts Dislocations Measure Topology in Graphene with Defects}% Force line breaks with \\

\author{Yuval Abulafia}
\author{Amit Goft}
\author{Nadav Orion}
\author{Eric Akkermans}%
\email{eric@physics.technion.ac.il}
\affiliation{%
Department of Physics, Technion -- Israel Institute of Technology, Haifa 3200003, Israel
}%

\date{\today}% It is always \today, today,

\begin{abstract}
We present a general method to identify topological materials by studying the local electronic density $\delta \rho \left(\boldsymbol{r}\right)$. More specifically, 
 certain types of defects or spatial textures  such as vacancies, turn graphene into a topological material characterised by invariant Chern or winding numbers. We show that these numbers are directly accessible from a dislocation pattern of $\delta \rho \left(\boldsymbol{r}\right)$, resulting from an interference effect induced by topological defects. For non topological defects such as adatoms, this pattern is scrambled by Friedel oscillations absent in topological cases. A Kekule distortion is discussed and shown to be  equivalent to a vacancy.

\end{abstract}

\pacs{Valid PACS appear here}% PACS, the Physics and Astronomy
\maketitle

%\tableofcontents

The identification and characterisation  of topological phases %\textcolor{magenta}{\cite{TKNN1982,KOHMOTO1985}} 
of matter \cite{Hasan2010,Qi2011} is a topic of recent and unabated interest. 
%There is an abundant literature which reflects the different points of view of how to describe topological features. 
The tenfold classification of insulators and superconductors proposed by Altland and Zirnbauer \cite{Altland1997} and extended to higher space dimensions %\citep{Kitaev2009,Schnyder2008}, 
provides a useful and elegant framework. It relies on %the existence of 
two anti-unitary symmetries, time reversal and particle-hole, the unitary chiral symmetry and space dimension $d$ of non-interacting systems of fermions  \citep{Kitaev2009,Schnyder2008}. For each symmetry class, relevant topology is identified by a  
homotopy group %based on the powerful yet abstruse $K$-theory classifying fibre bundles  
\citep{Kitaev2009,Stone2010}. The principle leading to this classification remains challenging, and it downplays the roles of defects and of disorder. These have been  considered in \cite{Teo2010,Chiu2016}, leading to a modified %generalisation of the 
tenfold classification and the replacement of $d$ by $\delta \equiv d - D$, where the dimension $D$ characterizes the spatial envelope of defects. An essential consequence of topologically non trivial classes is the appearance of gapless edge states  described  either by $\mathbb{Z}$ or $\mathbb{Z}_2$ integers, mainly accessible through transport or spectral measurements of edge states \cite{Aix2008,Hasan2010,Qi2011}. Topological features are encoded in the structure of wavefunctions rather than in the energy spectrum, which makes their observation challenging.

\begin{figure}[t]
\begin{centering}
\begin{tabular}{cc}
 {\includegraphics[width =0.5\columnwidth,height=4cm]{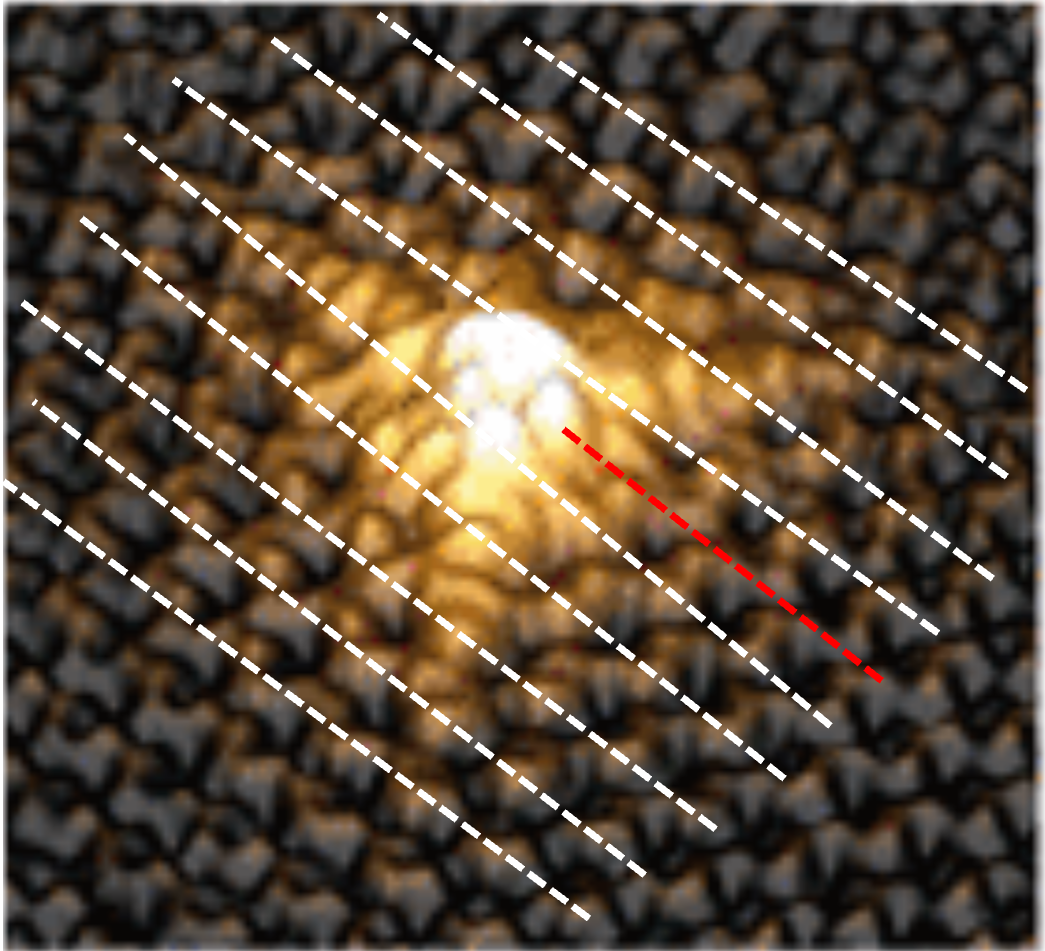}}&
 
 {\includegraphics[width =0.5
 \columnwidth,height=4cm]{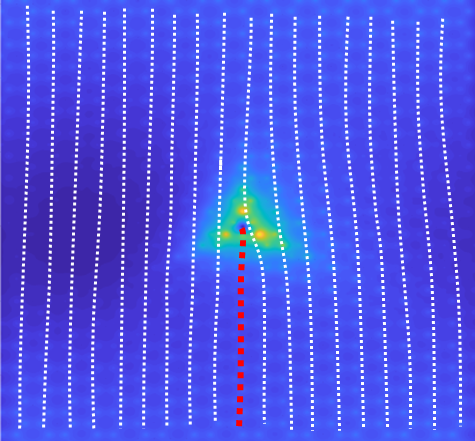}}
\end{tabular}
%\begin{subfigure}{0.33\textwidth}
 %   {\includegraphics[width =1\textwidth,height=4.5cm] {Kekule_n_minus2_BB.jpg}     \label{}}
%\end{subfigure}\\
\par\end{centering}
\centering{}\caption{Graphene with a vacancy. Left: STM measurement  \cite{Ugeda2010}, Right :  numerical result for $r^2 \delta\rho \left(\boldsymbol{r}\right)$ obtained for a $57 \times 60$ sites nearest neighbors tight binding model. %simulation (only $57 \times 43$ sites are displayed). %The hopping constant is $t=2.8 \, \text{eV}$. %and the lattice constant is $a=1$ 
In both images, white dashed lines were added to indicate wavefronts. The abruptly ending wavefront (red dashed line) corresponds to the dislocation of the electronic density.}
\label{fig:vacancy measurements and tight binding}
\end{figure}

In this paper we present a way to directly observe and measure $\mathbb{Z}$ topological integers by means of variations $\delta \rho \left(\boldsymbol{r}\right)$ of the local electronic density, a quantity directly accessible using scanning tunnel microscopy (STM). We show that  the existence of topological edge states leads to specific interference-induced dislocation patterns of $\delta \rho \left(\boldsymbol{r}\right)$  providing direct access to invariant integers, either Chern or winding. %(see Fig. \ref{fig:vacancy measurements and tight binding}). 
Our results apply to topological materials in general, yet we focus on two-dimensional systems and more specifically graphene, an always surprising material which provides a suitable playground to test these ideas. Pristine %and undoped 
graphene does not display %topological features nor 
protected edge states. It is no longer the case in the presence of specific types of defects, such as single atom vacancies or Kekule distortions which 
turn graphene into a topological material described by a  chiral symmetric Hamiltonian invariant under time reversal and particle-hole anti-unitary symmetries and with an effective dimension $\delta 
 =1$ \cite{goft2023}. Topology is characterised by an integer winding number $\nu_3 $ and zero energy edge states \cite{Teo2010,goft2023}. 
 These features  show up %are revealed 
 in a generic expression of the local density $\delta\rho \left(\boldsymbol{r}\right)$ induced by such defects or textures, 
 \begin{align}\label{local density vacancy first order}
\delta\rho \left(\boldsymbol{r}\right)  = F(r) 
 \, \sin \theta  \, \sin\left(\Delta\boldsymbol{K}\cdot\boldsymbol{r}+ N_D \, \theta\right)
\end{align}
where $\boldsymbol{r} = ( r, \theta)$ is measured from the location of the defect. The radial function $F (r)$ vanishes at infinity 
and $\Delta \boldsymbol{K} \equiv \boldsymbol{K}-\boldsymbol{K}'$ relates  the two independent Dirac points, i.e. the two valleys in graphene.  
The electronic density (\ref{local density vacancy first order}) vanishes at a point as displayed in Fig.\ref{fig:vacancy measurements and tight binding},  thus creating $ N_D$  dislocations, with $N_D =1$ for a single vacancy. The underlying mechanism
is presented in Fig.2. Using $\delta\rho \left(\boldsymbol{r}\right) = - \mbox{Re} \left( i F(r) \sin \theta \,  e^{i \chi (\boldsymbol{r})} \right)$ with 
\begin{equation}\label{phase}
 \chi (\boldsymbol{r}, N_D) \equiv \Delta\boldsymbol{K}\cdot\boldsymbol{r}+ N_D \, \theta  \, ,  
\end{equation}  allows to identify wavefronts i.e. planes of equal phase indicated by dotted lines in Fig.\ref{fig:vacancy measurements and tight binding}, and to rewrite (\ref{local density vacancy first order}) as,
\begin{equation} \label{interf}
\delta\rho \left(\boldsymbol{r}\right)  = F(r) \left[ \cos  \chi (\boldsymbol{r}, N_D -1 )   -  \cos  \chi (\boldsymbol{r}, N_D +1 ) \right].
\end{equation} 
Each %two 
oscillating term accounts from one %$\cos  \chi (\boldsymbol{r}, N_D -1 )$ and $\cos  \chi (\boldsymbol{r}, N_D +1 ) $ respectively originate from  
 sublattice (either $A$ or $B$), so that (\ref{interf}) describes a spatial interference pattern  between these two sublattices at a wave vector $\Delta \boldsymbol{K} $. This  emphasizes the role of inter-valley coupling induced by defects or textures. %, to observe interference between two sublattices electronic excitations. 
  A destructive interference %thus 
  corresponds to a dislocation, i.e. to the abrupt ending of a wavefront and to a vanishing of $\delta\rho \left(\boldsymbol{r}\right)$ \cite{Berry1980,NYE1974}.  
An important consequence of this interference-induced  dislocation is the existence of a winding number $\nu_3$ defined %obtained 
from the phase $\chi (\boldsymbol{r}, N_D) $, %namely
\begin{equation} \label{winding}
  \nu_3 \equiv \frac{1}{2\pi}\oint_{\cal C} \nabla\chi\left(\boldsymbol{r}\right)\cdot d\boldsymbol{r} = N_D   
\end{equation}
for any contour $\cal C$ encircling the defect location. This winding %number 
counts the number $N_D$ of dislocations, namely it characterises the interference pattern. We show that $\nu_3$ is the invariant integer expected from the tenfold classification for the BDI symmetry class in an effective dimension $\delta = 1$, i.e. for graphene with a vacancy or a Kekule distortion. Generally, observing %As a consequence, counting the number $N_D$ of 
dislocations in $\delta\rho \left(\boldsymbol{r}\right)$ allows direct access to invariant topological integers.%, here $\nu_3$. 
%This is a strong result which requires further proofs and  precautions in its use and understanding as explained below.  

The remainder of this letter is devoted to a derivation of these results,  to a discussion of the interference picture and its relation to invariant integers %topological numbers 
defined from the tenfold classification. We will consider different types %other examples 
of defects %either topological or not 
and decipher the corresponding %behavior of the 
change of electronic density $\delta\rho  \left(\boldsymbol{r}\right)$ so as to detect %identify 
topological features.% when applicable. 
We will conclude by explaining why and under which conditions, dislocation patterns in the electronic density provide a direct and indisputable way to identify and image %and measure topological numbers hence identify 
topological materials.  
%================ Begin Figure 2 ===========
%================ Ending Figure 2 ===========
%================ Begin Figure 1 ===========
\begin{figure}[t]
\begin{centering}
\begin{tabular}{cc}
 {\includegraphics[width =1\columnwidth,height=4cm]{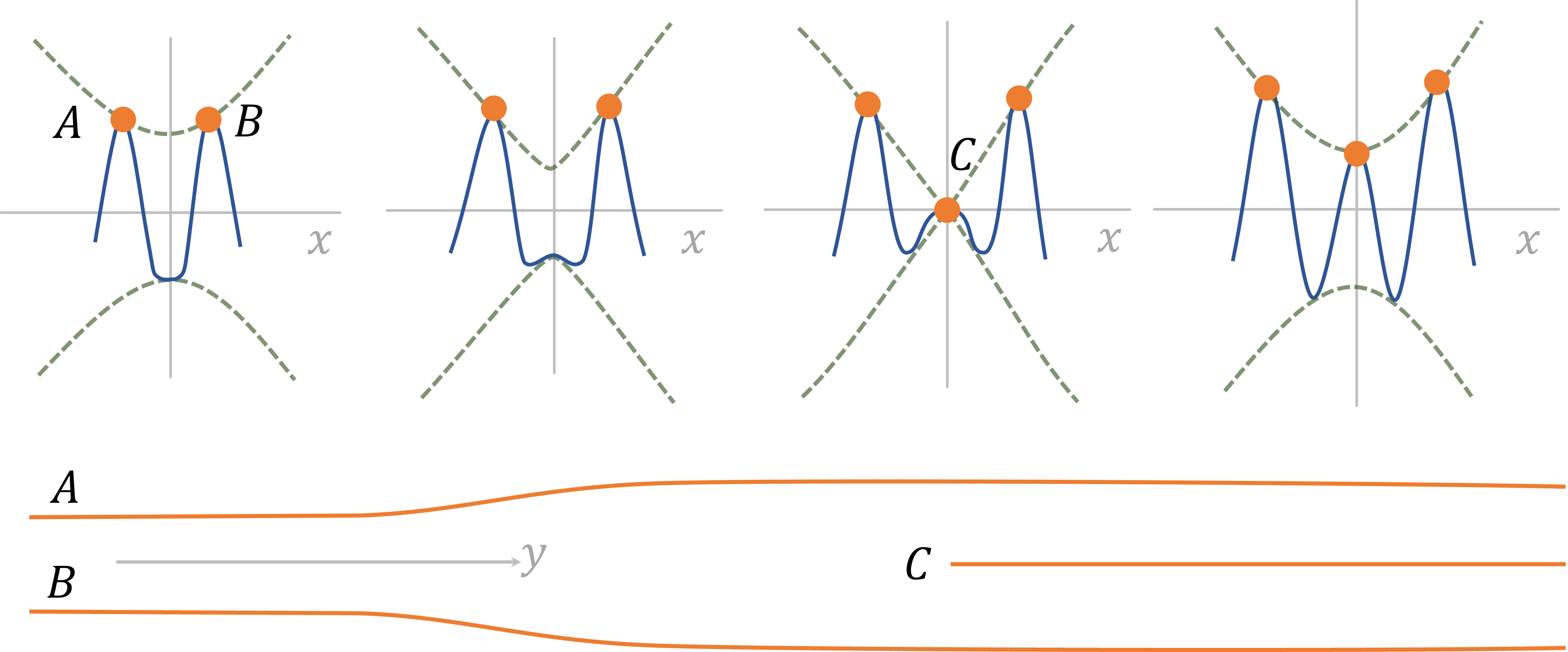}}&
\end{tabular}
%\begin{subfigure}{0.33\textwidth}
 %   {\includegraphics[width =1\textwidth,height=4.5cm] {Kekule_n_minus2_BB.jpg}     \label{}}
%\end{subfigure}\\
\par\end{centering}
%\begin{justifying}
\caption{Formation of a dislocation. (Top figures) A wave composed of an envelope (dashed green) multiplied by an oscillatory term (blue) presents  maxima (orange dots). Figures are plotted for increasing $y$ from left to right (continuous orange line in the bottom figure) while continuously tracking maxima $A$ and $B$. At points where  the amplitude vanishes, a new maximum $C$ appears between $A$ and $B$ and the corresponding line $C$ is a dislocation.}
%\end{justifying}
\label{fig:Dislocation_formation}
\end{figure}
%================ Ending Figure 1 ===========

We now establish expression (\ref{local density vacancy first order}) for undoped graphene with a vacancy, the simplest case of a topological defect. Graphene, a bipartite honeycomb lattice of carbon atoms, is appropriately  described by the tight-binding Hamiltonian, 
\begin{equation}\label{eq:tight_binding_H_0}
H_{0}=-t\sum_{\substack{i=0,1,2\\ \boldsymbol{R}\in A}}a_{\boldsymbol{R}}^{\dagger}b_{\boldsymbol{R}+\boldsymbol{\delta}_{i}}^{\phantom{\ast}}+h.c. 
\end{equation}
where $t$ is the hopping energy between nearest neighbor atoms belonging to two complementary triangular sublattices $A$ and $B$ with corresponding creation and annihilation operators  $(a_{\boldsymbol{R}}^{\dagger}$, $b_{\boldsymbol{R}}^{\vphantom{\ast}})$. 
The spectrum of $H_0$ displays two independent Dirac points, $K, K^\prime$ in the Brillouin zone. A vacancy
\cite{Kelly1998} is created by the removal of a neutral atom and of the bonds connecting
that atomic site to its neighbors. For a type $A$ vacancy at site
$\boldsymbol{R}_{0}$, the corresponding potential is \cite{Pereira2006,Peres2009,Peres2009A}, 
\begin{equation}
\hat{V}_{\boldsymbol{R}_{0}} \equiv+ \, t\sum_{i=1,2,3}a_{\boldsymbol{R}_{0}}^{\dagger}b_{\boldsymbol{R}_{0}+\boldsymbol{\delta_{i}}}^{\vphantom{\ast}}+h.c. 
\label{VA}
\end{equation}
where $\boldsymbol{\delta}_{i}$'s connect a site to its neighbor, with  $\boldsymbol{\delta}_{0}= \boldsymbol{0}$, $\boldsymbol{\delta}_{1}=-\boldsymbol{a}_{1}+\boldsymbol{a}_{2}$ and $\boldsymbol{\delta}_{2}=\boldsymbol{a}_{2}$, using the lattice vectors $\boldsymbol{a}_{1}=\sqrt{3}a \, \boldsymbol{\hat{x}}$ and $\boldsymbol{a}_{2}=\frac{\sqrt{3}}{2}a \, \boldsymbol{\hat{x}}+\frac{3}{2}a \, \boldsymbol{\hat{y}} $ and the lattice constant $a$.
We note that a vacancy does not add a new energy scale to the problem. %, and that 
$\hat{V}_{\boldsymbol{R}_{0}}$ involves only terms connecting sites $A$ to $B$  and, being %. Since it is 
a local potential, its Fourier transform includes high momentum terms leading to inter-valley scattering ($K \rightarrow K'$). A Fourier decomposition of $(a_{\boldsymbol{R}}^{\dagger}$, $b_{\boldsymbol{R}}^{\vphantom{\ast}})$ operators leads to,
%In the valley diagonal basis, $\psi =\left(\begin{array}{cccc}
%\psi_{A}^{K} & \psi_{B}^{K} & \psi_{A}^{K'} & \psi_{B}^{K'}\end{array}\right)^{T}$, it  is  a $4 \times 4$ operator valued matrix, 
\begin{align} \label{VAk}
\hat{V}_{\boldsymbol{R}_{0}} & =\frac{t}{N}\sum_{\boldsymbol{k},\boldsymbol{k'}}e^{i\left(\boldsymbol{k}-\boldsymbol{k}'\right)\cdot\boldsymbol{R}_{0}} f_{\boldsymbol{k'}} a_{\boldsymbol{k}}^{\dagger}b_{\boldsymbol{k}'}^{\vphantom{\ast}}+h.c.
\end{align}
with $f_{\boldsymbol{k'}} \equiv 1+e^{-i\boldsymbol{k'}\cdot\boldsymbol{\delta_{1}}}+e^{-i\boldsymbol{k'}\cdot \boldsymbol{\delta_{2}}}$. We split each sum over the  
Brillouin zone, i.e. $\sum_{\boldsymbol{k}} \equiv \sum_{\boldsymbol{k}}^{K}+\sum_{\boldsymbol{k}}^{K'}$, so as to identify two intra-valley $\left( \sum_{\boldsymbol{k},\boldsymbol{k'}}^{KK}\right)$ and two inter-valley $\left(\sum_{\boldsymbol{k},\boldsymbol{k'}}^{KK'}\right)$ terms. Despite similarities, they  behave  differently \cite{Mariani2007,Cheianov2006}, e.g. only inter-valley terms contribute to zero energy modes and to a winding number \cite{goft2023}. Keeping only them allows to rewrite (\ref{VAk}) under the form   $\hat{V}_{\boldsymbol{R}_{0}} =a^2 v_{F} \int d^{2} \boldsymbol{r} \, \psi_{\boldsymbol{r}}^{\dagger}\mathcal{\tilde V} \, \psi_{\boldsymbol{r}}^{\vphantom{\ast}}$ and to identify the operator, 
\begin{equation}
\mathcal{\tilde V}= \begin{psmallmatrix}
0 & 0 & 0 & -\delta\left(\boldsymbol{r-R_{0}}\right)L_{\textbf{r}}^{\dagger}\\
0 & 0 & \delta\left(\boldsymbol{r-R_{0}}\right)L_\textbf{r} & 0\\
0 & L_{\textbf{r}}^{\dagger}\delta\left(\boldsymbol{r-R_{0}}\right) & 0 & 0\\
-L_{\textbf{r}}\delta\left(\boldsymbol{r-R_{0}}\right) & 0 & 0 & 0
\end{psmallmatrix} \label{calV}
\end{equation}
written in the sublattice basis $ \left( \begin{matrix}
\psi_{A}^{K} & \psi_{A}^{K'} & \psi_{B}^{K} & \psi_{B}^{K'}\end{matrix} \right)^{T}$. We have defined $L_{\textbf{r}} \equiv -i\partial_x-\partial_y$ (SM section I) and $v_{F}=\frac{3}{2}ta$ is the Fermi velocity. We note that $\mathcal{\tilde V}$ clearly couples the two valleys $K$ and $K'$ and  preserves chiral symmetry. We anticipate an essential difference with the potential created by an adatom which breaks that symmetry  \cite{Yndurain2014,Nanda2013,Dutreix2016}.

The variation $\delta\rho_{A} \left(\boldsymbol{r}\right)$ of local density resulting from the creation of a type $A$ vacancy, is obtained from the standard expression of the energy integrated local density of states,
\begin{equation}\label{eq:local density and local density of states}
  \delta\rho_A \left(\boldsymbol{r}\right)= -\frac{1}{\pi} \int d\epsilon \, \text{Im} \, \delta G_\epsilon ^{R}\left(\boldsymbol{r},\boldsymbol{r} \right)   
\end{equation}
written using the retarded Green's function $G_\epsilon ^{R}$. The variation $\delta G_\epsilon ^{R}$ is expressed as a formal series expansion of the vacancy potential $\hat{V}_{\boldsymbol{R}_{0}}$. The first term of this expansion, $\delta\rho_A ^{(1)} \left(\boldsymbol{r}\right)$, is obtained from a product of $4 \times 4$ matrices,
\begin{equation}\label{eq:Green function to first order definition}
 \delta G_\epsilon ^{R (1)}\left(\boldsymbol{r},\boldsymbol{r}\right)=\int d\boldsymbol{r}_{1}d\boldsymbol{r}_{2}G_{0}^{R}\left(\boldsymbol{r},\boldsymbol{r}_{1}\right)V_A\left(\boldsymbol{r}_{1},\boldsymbol{r}_{2}\right)G_{0}^{R}\left(\boldsymbol{r}_{2},\boldsymbol{r}\right)   
\end{equation}
%deferring to later the study of the whole series. Expression (\ref{eq:Green function to first order definition}) 
where $G_{0}^{R}\left(\boldsymbol{r},\boldsymbol{r}_{1}\right) = \mbox{diag} \{ 
G_{0}^{K}, G_{0}^{K'} \}$ is graphene unperturbed Green's function in the valley diagonal basis. %$\left( \begin{matrix}
%\psi_{A}^{K} & \psi_{B}^{K} & \psi_{A}^{K'} & \psi_{B}^{K'}\end{matrix}\right)^T$ 
The valley Green's function is the $2\times2$ matrix, 
\begin{equation} \label{eq:graphene Green function}
G_{0}^{K}\left(\boldsymbol{r},\epsilon\right) \equiv  e^{i\boldsymbol{K}\cdot\boldsymbol{r}}
\begin{pmatrix}
G_{AA}^K & G_{AB}^{K}\\
G_{BA}^K & G_{BB}^K
\end{pmatrix}
\end{equation}
expressed in the sublattice diagonal basis, $G_{AA}^K = G_{BB}^K \equiv \frac{\epsilon}{v_F}\tilde{g}_\epsilon \left(r\right)$, $G_{AB}^{K} =  L_{\textbf{r}} \, \tilde{g}_\epsilon \left(r\right)$ and $\tilde{g}_\epsilon (r) \equiv \frac{- i}{4 v_{F}}H_{0}^{\left(1\right)}\left(\frac{\epsilon r}{v_{F}}\right)$ involves  a Hankel function.  
In the sublattice diagonal basis, %$\begin{pmatrix}
%\psi_{A}^{K} & \psi_{A}^{K'} & \psi_{B}^{K} & \psi_{B}^{K'}\end{pmatrix}^T$,  
the matrix element $V_A\left(\boldsymbol{r}_{1},\boldsymbol{r}_{2}\right) \equiv a^2  v_F  \langle \boldsymbol{r}_{1} | {\cal \tilde V} | \boldsymbol{r}_{2} \rangle$ is deduced from (\ref{calV}) (SM section III) and given by, 
\begin{equation}
\begin{psmallmatrix}
 &  &  & V_{AB}^{KK'}\\
 &  & V_{AB}^{K'K}\\
 & V_{BA}^{KK'}\\
V_{BA}^{K'K}
\end{psmallmatrix} \, .
\end{equation}
Overall, four non vanishing valley-coupling terms contribute to (\ref{eq:Green function to first order definition}) %$\delta G_\epsilon ^{R\, (1)}$
, denoted $\delta G_{AA}^{KK'}, \delta G_{BB}^{KK'}, \delta G_{AA}^{K'K}, \delta G_{BB}^{K'K}$ and symbolically written,
 \begin{equation}
\label{eq:first order valley and sublattice contributions}
\delta G_{AA}^{KK'} =\int \,  \left( G_{AA}^{K} V_{AB}^{KK'} G_{BA}^{K'} +G_{AB}^{K}V_{BA}^{KK'} G_{AA}^{K'} \right)  
\end{equation}
Summing over sublattices $A$ and $B$, we obtain two contributions, $\delta G_\epsilon ^{K'K}$ and %to the first order retarded Green's function, 
\begin{equation}
\delta G_\epsilon ^{KK'} =-2 a^2 \epsilon\, e^{i\Delta\boldsymbol{K}\cdot\boldsymbol{r}}\left[\tilde{g}_\epsilon \left(r\right)\nabla^{2} \tilde{g}_\epsilon \left(r\right)+\left(L_{\boldsymbol{r}}^{\dagger} \, \tilde{g}_\epsilon \left(r\right)\right)^{2}\right]
\end{equation}
%with 
%\begin{equation}
%\begin{cases}
%L_{\boldsymbol{x}}^{\dagger}L_{\boldsymbol{x}}=L_{\boldsymbol{x}}L_{\boldsymbol{x}}^{\dagger}=-\nabla^{2}\\
%L_{\boldsymbol{x}}g\left(\boldsymbol{r}\right)=-ie^{-i\theta}\frac{d}{dr}g\left(r\right)\\
%L_{\boldsymbol{x}}^{\dagger}g\left(\boldsymbol{r}\right)=-ie^{i\theta}\frac{d}{dr}g\left(r\right)
%\end{cases}
%\end{equation}
Inserting them into (\ref{eq:local density and local density of states}), an   integration over the valence band between a low energy cutoff $ - 3t$ and  $0$, leads to (\ref{local density vacancy first order}) with $N_D = 1$ and $F (r)  = {2 \sin \left( 4r/a \right) \over \pi^2 r^2} $  (SM section III). It is possible to generalise this calculation so as to include all higher order terms in the series (\ref{eq:local density and local density of states}), hence leading to (\ref{local density vacancy first order}) for $\delta \rho_A (\boldsymbol{r})$ (SM section IV) \footnote{In (\ref{calV}), we have neglected intra-valley terms which do not contribute to topological features. Yet, it must be stressed that to evaluate higher order terms, these omitted terms contribute but without changing the final result (\ref{local density vacancy first order}) up to a modification of the envelope function $F(r)$. }. 
  
We now consider another type of spatial defect created either by an adatom or by a substitutional atom in graphene. %, and compare it to the vacancy just studied. 
These potentials share several features with a vacancy, e.g. both have zero energy modes spatially located on the defect, but with distinct topological properties \citep{Yndurain2014,Nanda2013,Dutreix2016}. Unlike vacancies, creating an adatom \cite{Pereira2008, Palacios2008} involves a new  energy scale that breaks particle-hole and chiral symmetries, suggesting that an adatom belongs to $AI$ rather than $BDI$ class. Graphene with an adatom has a finite Fermi wave vector $k_F$ and the corresponding local electronic density $\delta \rho_{\text{ad}}\left(\boldsymbol{r}\right)$ displays, as expected, spatial Friedel oscillations at $2 k_F$ \cite{Bena2008,Dutreix2013,Dutreix2016}. A calculation \cite{Dutreix2019} based on (\ref{eq:local density and local density of states}) and (\ref{eq:Green function to first order definition}) leads to an expression similar to (\ref{interf}),   
\begin{equation} \label{interfad}
\delta \rho_{\text{ad}}^{\left(1\right)}\left(\boldsymbol{r}\right)  =f_A \left(k_F r\right)\cos\chi (\boldsymbol{r}, 0) -f_{B}\left(k_F r\right)\cos \chi (\boldsymbol{r}, 2)
\end{equation}
yet with essential differences. The two oscillating contributions of sublattices $A$ and $B$ show up, but they are weighted  by two radial and $2 k_F$-oscillating functions $f_{A,B} (k_F r)$ \cite{Dutreix2019} which differ for almost all $r$, $f_A \neq f_B$, hence destroying the interference and the dislocation pattern. As a result (\ref{interfad}) does not reduce to (\ref{interf}), and it does not exhibit a single dislocation $( N_D = 1)$ like for graphene with a vacancy. %This is a major difference between expressions (\ref{interf}) and (\ref{interfad}). 
While $\cos \chi (\boldsymbol{r}, 2)$ indeed leads to two dislocations at the location of the adatom \cite{Dutreix2019,Zhang2020,Zhang2021A} as displayed in Fig.\ref{Ringing_Friedel_for_adatom_and_vacancy}, these dislocations do not extend to infinity since the two $2 k_F$-oscillating functions $f_{A,B} (k_F r)$ disrupt the interference pattern each time they vanish. Hence, we observe an overall succession of areas with either two or zero dislocations. It is also clear from (\ref{interfad}) that a winding number cannot be calculated since the resulting value depends on the choice of the contour ${\cal C}$. We thus conclude that an adatom potential, unlike a vacancy, is not a topological defect \cite{goft2023}. These points were also discussed in one-dimensional setups \cite{Dutreix2017,Dutreix2021}.

\begin{center}\begin{figure}[ht]
\includegraphics[ width=1\linewidth]{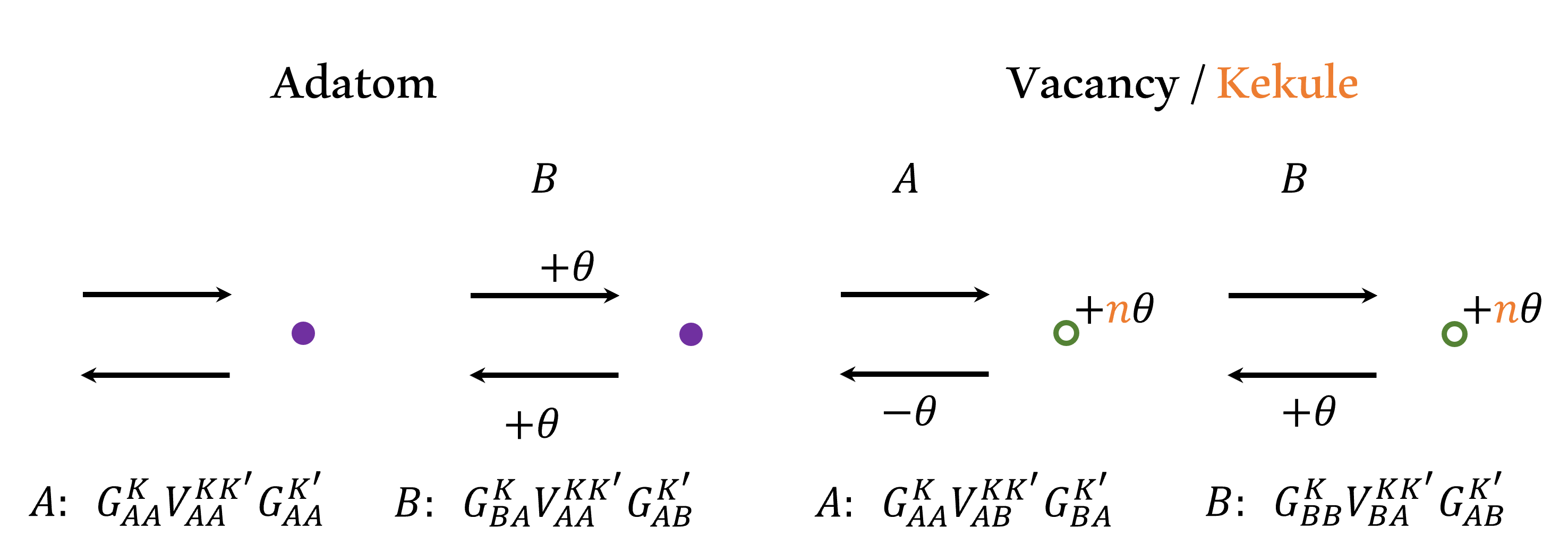}\caption{Interference and $\theta$-phase accumulation for scattering by a vacancy, a Kekule distortion or an adatom. Left: two scattering processes from an adatom. Sublattice $A$ contribution  does not accumulate phase whereas sublattice $B$ contribution accumulates $2\theta$ phase from unperturbed pristine graphene propagators. For a type $A$ vacancy (right), the phase accumulation in $A$ sublattice  cancels out whereas sublattice $B$ accumulates a $2\theta$ phase from the vacancy itself and pristine graphene. For a Kekule distortion, the two sublattices phase accumulations are respectively $ ( n \pm 1 ) \theta$.}
\label{perturb}
\end{figure}
\end{center}

\begin{figure}[t]
{\includegraphics[width =1\columnwidth,bb=0bp 0bp 520bp 545bp,clip]{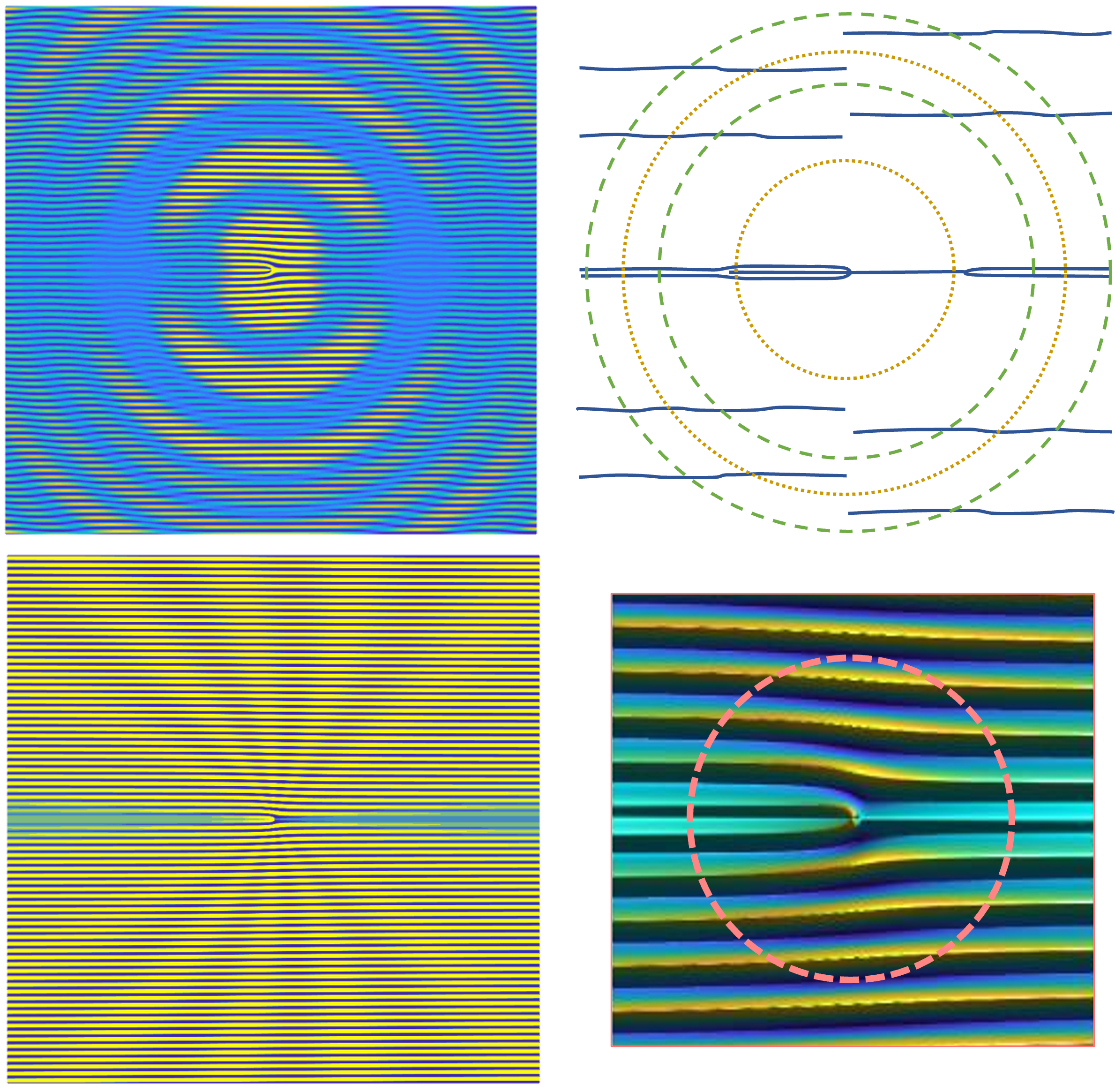}}
\caption{Plot of $r^2 \, \delta\rho \left(\boldsymbol{r}\right)$ for graphene with an adatom (top) and a vacancy (bottom). Top left: Spatial $2k_F $ radial Friedel oscillations for an adatom, plotted for %$V_0=50t$ and $V_b=0.33t$ so that 
$k_F \, a  \approx 0.22 $.  Plane wave oscillations at $|\Delta\boldsymbol{K}|^{-1} \approx a $ are along the vertical axis. %are represented in units of the lattice spacing $a$. 
Dislocations appear in different locations due to the $(2 k_F)^{-1}$  periodically vanishing radial amplitude. Top right: Track of the dislocations lines (in blue). The number of dislocations varies between 0 to 2 as function of the radius of the closed contour ${\cal C}$. Yellow (resp. green) dotted contours enclose two (resp. zero) dislocations. Bottom left: A single dislocation at the vacancy location extends to infinity irrespective of the chosen closed contour ${\cal C}$ (bottom right). }
\label{Ringing_Friedel_for_adatom_and_vacancy}
\end{figure}

It is interesting to uncover the origin of interference terms in (\ref{interf}) %and (\ref{interfad}) 
from the formal Green's function perturbation. We split contributions to (\ref{eq:Green function to first order definition}) into $\delta G_{AA}^{KK'}$  and $\delta G_{BB}^{KK'}$ from sublattices $A$ and $B$ respectively and their complex conjugates exchanging $K$ and $K'$. As indicated in Fig.\ref{perturb}, the two contributions in (\ref{eq:first order valley and sublattice contributions}), involve products of three terms : $G_{AA} ^K$ does not depend on $\theta$ while $V_{AB}^{KK'} G_{BA}^{K'}$ and $G_{AB}^{K} V_{BA}^{KK'}$ are respectively proportional to $e^{ \pm i (\theta - \theta) }$ namely are also $\theta$-independent, whereas sublattice $B$ contributions involve $e^{ \pm i (\theta + \theta) }$ terms. As a result for a type $A$ vacancy, sublattice $A$ (resp. $B$) terms contribute as $\cos \chi (\boldsymbol{r}, 0)$ (resp. as $\cos \chi (\boldsymbol{r}, 2)$) to (\ref{interf}). The net $\theta$-phase accumulation is the same for an adatom. Unlike a vacancy, the phase change  comes only from the unperturbed graphene Green's functions $G_{BA}^{K}$ and $G_{AB}^{K'}$, without influence from the adatom potential.% which does not participate to the interference process. 
 Moreover, the breaking of chiral symmetry leads to different radial functions, preventing an interference pattern.

%A further comparison between the two cases shows that for the adatom case the accumulation of $2\theta $ angle in the scattering process origins from pristine graphene propagation only, and not due to the adatom. This is not the case for a vacancy, whose by itself contributes an angle of $\pm \theta$ to the scattering process.

%================ Figure 5 =================
\begin{figure}[ht]
    \centering
    \begin{tabular}{cc}
      {\includegraphics[width =0.5\columnwidth,bb=150bp 120bp 500bp 380bp,clip]{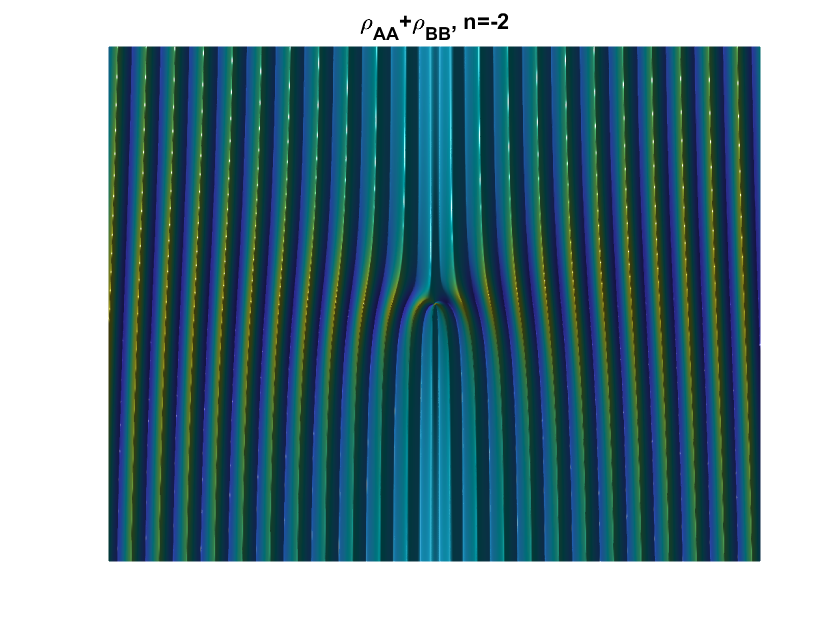}}
      &
\includegraphics[width =0.5\columnwidth,bb=150bp 120bp 500bp 380bp,clip]{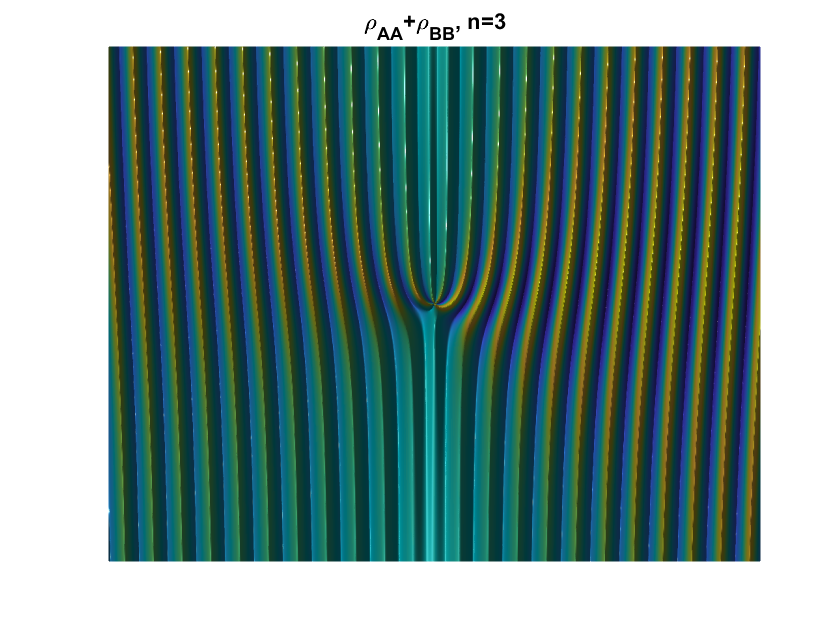}
      \\
      \small (a) & \small (b) 
      \\
	\end{tabular}
	\caption{Variation of the local density $r^2 \, \delta \rho_K ^{(1)} (\boldsymbol{r} )$ for graphene with a Kekule distortion with $\Delta_0\left(r\right)=-a\delta\left(r\right)$ and for different values of $n$. (a) $\nu_3 = 2$ dislocations for $n=-2$ and (b) $\nu_3 = 3$ dislocations of opposite direction for $n=3$. }
\label{fig:Kekule for different n's}
	\label{fig:dislocation formation}
\end{figure}

%================ Ending Figure 5 ============
A different example of valley coupling defect field in graphene, is the  Kekule distortion which recently came under scrutiny \cite{Hou2007,Jackiw2007,Gomes2012,Qu2022, Bao2021,Menssen2020}. Different versions exist which are essentially obtained by shifting the hopping term in (\ref{eq:tight_binding_H_0}) into $t_{\boldsymbol{r}, i}=t+\delta t_{\boldsymbol{r}, i}$ with 
\begin{equation}\label{kekule}
\delta t_{\boldsymbol{r},i}=\Delta_0\left(r\right)e^{i(n\theta + \alpha)} \exp{\left( i\boldsymbol{K}\cdot\boldsymbol{\delta}_{i} + i \, \Delta\boldsymbol{K}\cdot\boldsymbol{r} \right)}\, .  
\end{equation}
This specific type of Kekule distortion corresponds to a  centrosymmetric field $\Delta_0\left(r\right)e^{i(\alpha +n\theta)}$ where $\boldsymbol{r} = (r,\theta)$  is  measured from the origin and $\alpha$ is constant. This Kekule distortion and a vacancy are topologically equivalent defects \cite{goft2023} both characterised by a fractional charge \cite{Ovdat2017,Ovdat2020,Hou2007} and an integer winding number $\nu_3 = N_D = n$ given by (\ref{winding}) and revealed in the dislocation pattern of the local electronic density $\delta \rho_K (\boldsymbol{r} )$. More precisely, starting from for the Kekule model described in \cite{Hou2007} and using a rotation in the pseudo-spin space, we obtain a potential ${\cal V}_K$  similar to $ \cal \tilde V$ provided we replace $\delta\left(\boldsymbol{r}- \boldsymbol{R}_{0}\right)L_{\textbf{r}}^{\dagger}$ by $ \Delta_{0}\left(r\right)e^{-in\theta-i\alpha}$ and $L_{\textbf{r}}\delta\left(\boldsymbol{r}-\boldsymbol{R}_{0}\right)$ by $\Delta_{0}^{*}\left(r\right)e^{in\theta+i\alpha}$ (SM section V). Hence similar calculations (SM section VI) lead to (\ref{interf}) and (\ref{winding}) for the first order correction $\delta \rho_K ^{(1)} (\boldsymbol{r} )$. We then observe  $N_D = n$ dislocations and an interference pattern explained in Fig.\ref{perturb} and visible in Fig.\ref{fig:Kekule for different n's} for a sufficiently rapidly decreasing field $\Delta_{0}\left(r/a \right)$. 
%================ Figure 4 =================
\begin{figure}[t]

\end{figure}
%================ Ending Figure 4 ===========
%For graphene with a vacancy and with a Kekule distortion, two cases with chiral symmetry and $\mathbb{Z}$-topology, we have shown that the variation of local density $\delta \rho_K (\boldsymbol{r} )$ displays $N_D$ dislocations corresponding to a winding number $\nu_3$ given in (\ref{winding}). Following the tenfold classification, these two cases belong to the $\left( s, \delta \right) = (1,1)$ class whose $\mathbb{Z}$-valued topological invariants are winding numbers \cite{goft2023}. 

The existence of $\mathbb{Z}$ topological defects is related to the appearance of zero energy modes in the spectrum of the Hamiltonian \citep{Ryu2002, Brouwer2002, Ganeshan2013}. To obtain them for the vacancy potential $ \hat{V}_{\boldsymbol{R}_{0}}$ in (\ref{VA}), we look for solutions of $H  \psi=0$ with $H \equiv \left( H_0 + a^2 v_{F} \cal V \right)$. Expression  (\ref{calV}) leads to a single zero mode (SM section II). This result could have been anticipated on the basis of the formal yet powerful Atiyah-Singer index theorem \citep{Atiyah1963, Atiyah1968,Niemi&Semenoff1984} which relates the analytical index, i.e. the counting of zero modes of $H$ to the topological index calculated from its symbol \cite[Chapters~ 24-27]{yankowsky2013,goft2023}.  %For a vacancy, i.e. a chiral topological defect, there is a relation between the topological winding number $\nu_3$ in (\ref{winding}). 
A single vacancy or a Kekule distortion is described by chiral Hamiltonians of generic  form $ H= \begin{psmallmatrix}
        0 & \cal{Q} \\
        \cal{Q}^{\dagger} & 0 
    \end{psmallmatrix} $  
for which the winding number $\nu_3$ also counts zero energy edge states given by the analytical index, 
 $ \text{Index} \, H \equiv \text{dim}\, \text{Ker} \, \cal{Q}  -  \text{dim}\, \text{Ker} \, \cal{Q}^{\dagger} = \# \, \text{of zero modes}$. This analytical index equals the topological index, here the winding number \cite{goft2023} namely,  $\text{Index} \, H = \nu_3 $. Finally, using (\ref{winding}), leads to 
%we first note that $N_D$ is a characteristic of $\delta \rho \left(\boldsymbol{r}\right)$, a quantity obtained from the Hamiltonian $H_V$ of graphene with a vacancy, while the topological winding number is obtained from the symbol $\mathcal{H} (\boldsymbol{k}, \boldsymbol{r})$ of the Hamiltonian \cite{goft2023}. There is a relation between a Hamiltonian and its symbol. In the presence of  translation symmetry, the symbol is nothing but the Bloch Hamiltonian obtained from the Hamiltonian by a unitary transformation. This state of affairs is the exception rather than the rule and although Bloch theorem does not apply when translation symmetry is broken, e.g. in the presence of graphene with a vacancy, the symbol is well defined. It does no longer provide information on the Hamiltonian spectrum, but it encodes all topological information \citep{Atiyah1963,Atiyah1968, Nakahara1990}, \cite[Chapters~26-27]{yankowsky2013}. 
% Topologically invariant numbers can be retrieved from $\mathcal{H} (\boldsymbol{k}, \boldsymbol{r}) $ provided a spectral gap exists in its  eigenvalue spectrum. In that case, the Index theorem $\text{Index} \, H = \nu_3$ holds between the Hamiltonian and its symbol \cite{goft2023}, so that
 \begin{equation}
     N_D = \nu_3 = \text{Index} \, H = \# \, \text{of zero modes} \, .
 \end{equation}  
 This relation links the number $N_D$ of dislocations and the topological winding number $\nu_3$. It applies to more than one vacancy, e.g. for a dimer defined by the removal of two neighboring atoms from $A$ and $B$ sublattices, so that  $\text{Index} \, H = \nu_3 = 0$, namely the absence of dislocations $N_D =0$. 

 We now summarise our findings and discuss some of their consequences. We have shown that topological integers %numbers 
 defined in the framework of the tenfold classification are directly observable from a dislocation pattern of the local electronic density. Our results have been obtained for two-dimensional chiral systems, specifically for graphene with two types of defects: vacancy and Kekule distortion. The dislocation pattern is evidence from an underlying spatial interference between the two graphene sublattices, modulated at a wavevector $\Delta\boldsymbol{K}$ associated with the coupling between valleys. A %topological 
 winding number is directly retrieved from this interference which is scrambled for non topological defects like e.g. adatoms. Application of these results to other symmetry classes and spatial dimensions is promising. For instance, a material with both vacancies and adatoms breaks chiral symmetry and belongs to the non-chiral %topological 
 class $AI$ in an effective dimension $\delta =0$, hence  assigned a Chern instead of a winding number with a specific dislocation pattern. The creation of defects can turn %transform 
 a given material into topological \cite{goft2023}. Relevant information on such transitions between topological phase is accessible using the evolution of dislocation and interference patterns. Finally, another situation to unveil interesting topology by means of dislocations is in multilayered, e.g. Moire, materials where our approach may prove useful.

\section*{Acknowledgments}
\begin{acknowledgments}
This work was supported by the Israel Science Foundation Grant No.~772/21 and by the Pazy Foundation.
\end{acknowledgments}.

%---- References ------%

%\bibliography{Thesis.bib} 

%----------------------------------%
\end{document}

\title{Supplementary material for
"Measuring topological winding number by counting wavefront dislocations"}
\author{Yuval Abulafia, Amit Goft, Nadav Orion and Eric Akkermans}
\date{\today}
{
\let\clearpage\relax
\maketitle
}
\section{Modeling a Vacancy}
\renewcommand{\theequation}{S.\arabic{equation}}
\setcounter{equation}{0}
In this section, we establish expression (8) in the main text for the potential of a vacancy in the low energy limit. We start with the tight binding description of a vacancy as given by (5) and (6) and expand $a_{\boldsymbol{R}_{0}}^{\dagger}$, and $b_{\boldsymbol{R}_{0}+\boldsymbol{\delta_{i}}}^{\vphantom{\ast}}$ in momentum space as a sum over the Brillouin zone
 \begin{align}\label{eq:S1}
\hat{V}_{\boldsymbol{R}_{0}} & = \frac{t}{N}\sum_{\boldsymbol{k},\boldsymbol{k'}}e^{i\left(\boldsymbol{k}-\boldsymbol{k}'\right)\cdot\boldsymbol{R}_{0}}\left[1+e^{-i\boldsymbol{k}'\cdot\boldsymbol{\delta}_{1}}+e^{-i\boldsymbol{k}'\cdot\boldsymbol{\delta}_{2}}\right]a_{\boldsymbol{k}}^{\dagger}b_{\boldsymbol{k}'}^{\vphantom{\ast}}+h.c. \, ,
\end{align}
%In order to access the two valleys we shall 
which is split into  
$\sum_{\boldsymbol{k}} \equiv \sum_{\boldsymbol{k}}^{K}+\sum_{\boldsymbol{k}}^{K'}$.
%The two summations over the momenta lead to four terms, two intravalley $\sum_k^{K}\sum_{k'}^{K}$, $\sum_k^{K'}\sum_{k'}^{K'}$ and two inter-valleys $\sum_k^{K}\sum_{k'}^{K'}$, $\sum_k^{K'}\sum_{k'}^{K}$.
%Throughout the derivation, the inter-valley and intravalley terms are similar.
As mentioned in the main text, this decomposition distinguishes inter-valley and intra-valley contributions. %which couple different operators. 
%This translates to their location within $4\times4$ matrix of the vacancy potential.
%Intra-valley contributions appear in the same matrix slots as $H_0$, therefore do not create new coupling.
Unlike intra-valley, inter-valley interactions give rise to new couplings  between uncoupled operators of pristine graphene.
%As a result, even though their derivation is similar, inter-valley and intravalley terms contribution is drastically different. 
We neglect intra-valley terms in our calculations, as only inter-valley terms are responsible for zero energy solutions (see next section) and topological winding numbers \cite{Goft2023}. \\

For the two inter-valley terms $(KK')$, each summation indicates the point around which we expand wavevectors, namely %e.g. for $\sum_{\boldsymbol{k}}^{K}\sum_{\boldsymbol{k}'}^{K'}$, expand  
$\boldsymbol{k}=\boldsymbol{K}+\delta\boldsymbol{k}$ and $\boldsymbol{k}'=\boldsymbol{K'}+\delta\boldsymbol{k}'$, e.g.  $a_{\boldsymbol{k}}$ ($b_{\boldsymbol{k}'}$) operator,  around the $K$ ($K'$) Dirac point becomes $a_{\delta\boldsymbol{k}}^{K}=e^{-i \boldsymbol{K} \cdot \boldsymbol{R}_{0}} a_{\boldsymbol{k} =\boldsymbol{K} +\delta\boldsymbol{k}}$ (similarly for 
$b_{\delta\boldsymbol{k}'}^{K'}$). This step implies enlarging the  basis so as to include both sublattices and valleys. %in exchange for limiting the Brillouin zone region. 
A first order expansion of \eqref{eq:S1} in $\delta k'a$ leads to 
\begin{align}
\hat{V}_{\boldsymbol{R}_{0}} & =\frac{v_F}{N}\left(\sum_{\delta\boldsymbol{k}} ^{K} \sum_{\delta\boldsymbol{k}'} ^{K'}\right)e^{i\left(\delta\boldsymbol{k}-\delta\boldsymbol{k}'\right)\cdot\boldsymbol{R}_{0}}\left(\delta k_{x}'+i\delta k_{y}'\right)a_{\delta\boldsymbol{k}}^{K\dagger}b_{\delta\boldsymbol{k}'}^{K'}+h.c.\\
 & +\frac{v_F}{N}\left(\sum_{\delta\boldsymbol{k}}^{K'}\sum_{\delta\boldsymbol{k}'}^{K}\right)e^{i\left(\delta\boldsymbol{k}-\delta\boldsymbol{k}'\right)\cdot\boldsymbol{R}_{0}}\left(-\delta k_{x}'+i\delta k_{y}'\right)a_{\delta\boldsymbol{k}}^{K'\dagger}b_{\delta\boldsymbol{k}'}^{K}+h.c.\nonumber \, . 
\end{align}
The summation over $\delta\boldsymbol{k}'$ is performed using 
$\sum_{\delta\boldsymbol{k}'}^{K'}\delta k_{x}' \, e^{-i \, \delta\boldsymbol{k}'\cdot\boldsymbol{R}_{0}} \, b_{\delta\boldsymbol{k}'}^{K'}=i\partial_{x_{0}} \, b_{\boldsymbol{R}_0}^{K'}$, with obvious notations, so that 
\begin{align}
\hat{V}_{\boldsymbol{R}_{0}} & = v_F \, a_{\boldsymbol{R}_{0}}^{K\dagger}\left(i\partial_{x_{0}}-\partial_{y_{0}}\right)b_{\boldsymbol{R}_{0}}^{K'}+h.c.\nonumber \\
 & + v_F \, a_{\boldsymbol{R}_{0}}^{K'\dagger}\left(-i \, \partial_{x_{0}}-\partial_{y_{0}}\right)b_{\boldsymbol{R}_{0}}^{K}+h.c.
\end{align}
which rewrites
\begin{align}
\hat{V}_{\boldsymbol{R}_{0}} & = v_F \int d\boldsymbol{r} \, a_{\boldsymbol{r}}^{K\dagger}\delta\left(\boldsymbol{r}-\boldsymbol{R}_0\right)\left(i \, \partial_{x}-\partial_{y}\right)b_{\boldsymbol{r}}^{K'}+h.c.\nonumber \\
 & + v_F \, \int d \boldsymbol{r} \,  a_{\boldsymbol{r}}^{K'\dagger}\delta\left(\boldsymbol{r}-\boldsymbol{R}_0\right)\left(-i \, \partial_{x}-\partial_{y}\right)b_{\boldsymbol{r}}^{K}+h.c.
\end{align}
defining $L_{\boldsymbol{r}} \equiv -i \, \partial_x-\partial_y$
%and $L_{\boldsymbol{r}}^{\dagger}=-i\partial_x+\partial_y$
and using a matrix form, %including the hermitian constants
\begin{equation}
\hat{V}_{\boldsymbol{R}_{0}}=\int d\boldsymbol{r}\, \psi_{\boldsymbol{r}}^{\dagger}\, \mathcal{V}\, \psi_{\boldsymbol{r}}
\end{equation}
\begin{equation}
\mathcal{V}=a^2 v_{F}\left(\begin{array}{cccc}
0 & 0 & 0 & -\delta\left(\boldsymbol{r}-\boldsymbol{R}_0\right)L_{\boldsymbol{r}}^{\dagger}\\
0 & 0 & \delta\left(\boldsymbol{r}-\boldsymbol{R}_0\right)L_{\boldsymbol{r}} & 0\\
0 & L_{\boldsymbol{r}}^{\dagger}\delta\left(\boldsymbol{r}-\boldsymbol{R}_0\right) & 0 & 0\\
-L_{\boldsymbol{r}}\delta\left(\boldsymbol{r}-\boldsymbol{R}_0\right) & 0 & 0 & 0
\end{array}\right)
\end{equation}
where %we defined $v_{F}=\frac{3}{2}ta$ and 
$\psi_{\boldsymbol{r}}\equiv \left(\begin{array}{cccc}
\psi_{A}^{K}\left(\boldsymbol{r}\right) & \psi_{A}^{K'}\left(\boldsymbol{r}\right) & \psi_{B}^{K}\left(\boldsymbol{r}\right) & \psi_{B}^{K'}\left(\boldsymbol{r}\right)\end{array}\right)^{T}$ 
 written in the sublattice basis. Note that $\psi_{\boldsymbol{r}}$ has dimensions of inverse of a length and %has units of $1/a$, opposed to 
$a_{\boldsymbol{r}} ^K$, $b_{\boldsymbol{r}} ^K$ are dimensionless. This is relation (8) given in the text for ${\cal V} = a^2  v_F {\cal \tilde V}$.

\section{Graphene with a vacancy : zero energy solutions %and two dimensional angular momentum
}
In this section, we show that (8) (graphene with a vacancy) leads to a single zero energy solution (hereafter zero mode) and %use the two dimensional angular momentum to find 
calculate it. We write the Hamiltonian of graphene with a vacancy in first quantisation using the same notation as for the Hamiltonian of pristine and undoped graphene, $H_0$ as presented in (5) in the low energy limit
\begin{align}\label{eq:graphene with vacancy Hamitonian}
H_{\text{0}} + \mathcal{V} & =v_{F}\left(\begin{array}{cccc}
0 & 0 & L_{\boldsymbol{r}} & -a^2\delta\left(\boldsymbol{r}-\boldsymbol{R}_0\right)L_{\boldsymbol{r}}^{\dagger}\\
0 & 0 & a^2\delta\left(\boldsymbol{r}-\boldsymbol{R}_0\right)L_{\boldsymbol{r}} & -L_{\boldsymbol{r}}^{\dagger}\\
L_{\boldsymbol{r}}^{\dagger} & a^2L_{\boldsymbol{r}}^{\dagger}\delta\left(\boldsymbol{r}-\boldsymbol{R}_0\right) & 0 & 0\\
-a^2 L_{\boldsymbol{r}}\delta\left(\boldsymbol{r}-\boldsymbol{R}_0\right) & -L_{\boldsymbol{r}} & 0 & 0
\end{array}\right)\, ,
\end{align}
and solve $\left( H_{\text{0}} + \mathcal{V} \right)\psi=0$ for zero modes. Without loss of generality, we take the vacancy location at the origin $\boldsymbol{R}_{0} = \boldsymbol{0}$. We define the Dirac-$\delta$ functions as a limit of some real valued and analytical function that vanishes for $r>r_0$ %, and use the eigenfunctions in this area, where the vacancy potential vanishes
. The solutions at $r>r_0$ constitute a set of angular modes $\psi\left(\boldsymbol{r}\right)= \sum_{m=-\infty}^{\infty}e^{im\theta}\psi_{m}\left(r\right)
$ where the zero modes equation requires $\psi_{A,m}^{K}\left(r\right)$ and $\psi_{B,m}^{K'}\left(r\right)$ to behave like $r^m$ while  $\psi_{A,m}^{K'}\left(r\right)$ and $\psi_{B,m}^{K}\left(r\right)$ to behave like $r^{-m}$. Not all modes show up since the wavefunction must vanish at infinity, which leads to truncating modes in each sum, 
\begin{align}\label{eq:vanishing_at_infty_modes}
\psi_{A}^{K}\left(\boldsymbol{r}\right)=\sum_{m=-\infty}^{-1}A_{m}^{K}e^{im\theta}r^{m} & ,\: & \psi_{B}^{K}\left(\boldsymbol{r}\right)=\sum_{m=1}^{\infty}B_{m}^{K}e^{im\theta}r^{-m}\\
\psi_{A}^{K'}\left(\boldsymbol{r}\right)=\sum_{m=1}^{\infty}A_{m}^{K'}e^{im\theta}r^{-m} & ,\: & \psi_{B}^{K'}\left(\boldsymbol{r}\right)=\sum_{m=-\infty}^{-1}B_{m}^{K'}e^{im\theta}r^{m}\nonumber
\end{align}
We introduce the 2-dimensional angular momentum algebra, defining the operator $L\equiv -i \, \left(x \, \partial_y - y \, \partial_x\right)$ which abides $L\left|m\right\rangle= m\left|m\right\rangle$ where $\left\langle\boldsymbol{r}|m\right\rangle\propto e^{im\theta}$. We then  calculate its commutation relation with  $L_{\boldsymbol{r}}$ and $L_{\boldsymbol{r}}^{\dagger}$ following a ladder behavior, 
\begin{align}\label{eq:ladder_op_def}
    L_{\boldsymbol{r}}\left|m\right\rangle&=\left|m-1\right\rangle\\
    L_{\boldsymbol{r}}^{\dagger}\left|m\right\rangle&=\left|m+1\right\rangle.\nonumber
\end{align}
Let us write $\left(H_0+\cal V\right)\psi=0$ explicitly,  
\begin{align}
 & L_{\boldsymbol{r}}^{\dagger}\psi_{A}^{K}+a^2L_{\boldsymbol{r}}^{\dagger}\delta\left(\boldsymbol{r}\right)\psi_{A}^{K'}=0\\
 & L_{\boldsymbol{r}}\psi_{A}^{K'}+a^2L_{\boldsymbol{r}}\delta\left(\boldsymbol{r}\right)\psi_{A}^{K}=0\nonumber\\
 & L_{\boldsymbol{r}}\psi^K_B-a^2\delta\left(\boldsymbol{r}\right)L_{\boldsymbol{r}}^\dagger\psi_{B}^{K'}=0\\
 &L_{\boldsymbol{r}}^{\dagger}\psi_{B}^{K'}-a^2\delta\left(\boldsymbol{r}\right)L_{\boldsymbol{r}}\psi_{B}^{K}=0\nonumber
\end{align}
we insert \eqref{eq:vanishing_at_infty_modes} and use the ladder operators \eqref{eq:ladder_op_def} to observe which of the modes are coupled (summarised in Table \ref{Table:modes}). On sublattice $A$, each mode $m$ of  $\psi_{A,m}^K$ is coupled to the same mode $m$ of $\psi_{A,m}^{K'}$. Since from \eqref{eq:vanishing_at_infty_modes} the two valleys share no angular modes (a $"0"$ in Table \ref{Table:modes}), sublattice $A$ does not support zero energy  solutions. However, on sublattice $B$ there is only one option to couple two modes (marked in orange in Table \ref{Table:modes}), thus providing a single zero energy solution. Since the two equations (S.12) relative to sublattice $B$, are symmetrical and the delta function is real valued, the valley coefficients of the pseudospinor must be equal. Writing the single solution while considering the shifted momentum of each valley provides,
\begin{table}
\small
\begin{center}\begin{tabular}{|c|c|c|c|c|}
\hline 
$m$ & $\psi_A^K$ & $\psi_A^{K'}$ & $\psi_B^{K}$ & $\psi_B^{K'}$ 
\\
\hline 
\hline 
 $\vdots$ & 0 & \cellcolor{lightgray} & \cellcolor{lightgray} &0\\
 \hline
3 & 0 & \cellcolor{lightgray} & \cellcolor{lightgray} & 0
\\
\hline 
2 & 0 & \cellcolor{lightgray} & \cellcolor{lightgray} & 0
\\ 
\hline
1 & 0 & \cellcolor{lightgray} & \cellcolor{orange} & 0
\\
\hline 
0 & 0 & 0 & 0 & 0
\\
\hline 
 -1 & \cellcolor{lightgray} & 0 & 0 & \cellcolor{orange}
\\
\hline 
  -2 & \cellcolor{lightgray} & 0 & 0& \cellcolor{lightgray}
\\
\hline 
 -3& \cellcolor{lightgray} & 0 & 0 & \cellcolor{lightgray}
\\
\hline 
 $\vdots$ & \cellcolor{lightgray} & 0 & 0 &\cellcolor{lightgray}

\\
\hline
\end{tabular}
\end{center}
\caption{Angular modes $m$ for each sublattice and valley are listed in the left column. Vanishing (resp. non-vanishing) modes associated to the behavior at $r\to\infty$ are marked by $0$ (resp. grey slots). Among all modes, only two non-vanishing modes (in orange) are coupled by the vacancy potential at $r=0$. All other grey modes vanish since they stay uncoupled at $r=0$.   \label{Table:modes}}
\end{table}
\begin{equation}
\psi_{\boldsymbol{r}} \propto \frac{1}{r}\left(\begin{array}{c}
0\\
0\\
e^{iK\cdot r+i\theta}\\
e^{iK'\cdot r-i\theta}
\end{array}\right) \, ,
\end{equation}
where we emphasize that this zero energy solution is not normalizable.

%%%%%%%%%%%%%%%%%%%%%%%%%%%%%%%%%%
%%%%%%%%%%%%%%%%%%%%%%%%%%%%%%%%%%%%%
%%%%%%%%%%%%%%%%%%%%%%%%%%%%%%%%%%%%%%%%%%%%%%%%%%%

%%% ======================================================================================================================================================================================================================================================================================== %%% 
\section{Local density for a single vacancy}
In this section, we establish expression (14) of the main text. We start by  writing explicitly the Green's functions for pristine graphene in the two valleys
\begin{align}\label{eq:graphene Green function}
G_{0}^{K}\left(\boldsymbol{r},\epsilon\right)& = e^{i\boldsymbol{K}\cdot\boldsymbol{r}}\left(\begin{array}{cc}
\frac{\epsilon}{v_{F}}\tilde{g}_{\epsilon}\left(r\right) & \left(-i\frac{d}{dx}-\frac{d}{dy}\right)\tilde{g}_{\epsilon}\left(r\right)\\
\left(-i\frac{d}{dx}+\frac{d}{dy}\right)\tilde{g}_{\epsilon}\left(r\right) & \frac{\epsilon}{v_{F}}\tilde{g}_{\epsilon}\left(r\right)
\end{array}\right) \equiv\left(\begin{array}{cc}
G_{AA}^K & G_{AB}^{K}\\
G_{BA}^K & G_{BB}^K
\end{array}\right) \nonumber\\  
G_{0}^{K'}\left(\boldsymbol{r},\epsilon\right)& = e^{i\boldsymbol{K}'\cdot\boldsymbol{r}}\left(\begin{array}{cc}
\frac{\epsilon}{v_{F}}\tilde{g}_{\epsilon}\left(r\right) & -\left(-i\frac{d}{dx}+\frac{d}{dy}\right)\tilde{g}_{\epsilon}\left(r\right)\\
-\left(-i\frac{d}{dx}-\frac{d}{dy}\right)\tilde{g}_{\epsilon}\left(r\right) & \frac{\epsilon}{v_{F}}\tilde{g}_{\epsilon}\left(r\right)
\end{array}\right)
\end{align}
where $\tilde{g}_{\epsilon}\left(r\right)= - \frac{i}{4v_{F}}H^{\left(1\right)}_{0}\left(\frac{\epsilon r}{v_{F}}\right)$ \cite{DutreixThesis,Bena2008} is written using the Hankel function (Bessel function of the third kind). %\textcolor{magenta}{We redefine $g_{\epsilon}\left(\boldsymbol{r}\right)=\frac{\epsilon}{v_{F}}\tilde{g}_{\epsilon}\left(\boldsymbol{r}\right)$}

For the vacancy potential, as seen in (12) of the main text, the non-vanishing terms alternate valley as well as sublattice. Starting from equation (9) in the main text, we explicitly write all contributions  as was done symbolically in expression (13) in the main text and using the notation $\left\langle\boldsymbol{r}\right|\delta G_{AA}^{KK'} \left|\boldsymbol{r}'\right\rangle=\delta G_{AA}^{KK'} \left(\boldsymbol{r},\boldsymbol{r'}\right)$, lead to the set of equations,
 \begin{align}\label{eq:first order valley and sublattice contributions}
\delta G_{AA}^{KK'}\left(\boldsymbol{r},\boldsymbol{r}'\right)  & =\int d\boldsymbol{r}_1d\boldsymbol{r}_2G_{AA}^{K}\left(\boldsymbol{r}-\boldsymbol{r}_1\right)V_{AB}^{KK'}\left(\boldsymbol{r}_1,\boldsymbol{r}_2\right)G_{BA}^{K'}\left(\boldsymbol{r}_2-\boldsymbol{r}'\right)\\
 & +\int d\boldsymbol{r}_1d\boldsymbol{r}_2G_{AB}^{K}\left(\boldsymbol{r}-\boldsymbol{r}_1\right)V_{BA}^{KK'}\left(\boldsymbol{r}_1,\boldsymbol{r}_2\right)G_{AA}^{K'}\left(\boldsymbol{r}_2-\boldsymbol{r}'\right)\nonumber\\
 \delta G_{AA}^{K'K}\left(\boldsymbol{r},\boldsymbol{r}'\right)  & =\int d\boldsymbol{r}_1d\boldsymbol{r}_2G_{AA}^{K'}\left(\boldsymbol{r}-\boldsymbol{r}_1\right)V_{AB}^{K'K}\left(\boldsymbol{r}_1,\boldsymbol{r}_2\right)G_{BA}^{K}\left(\boldsymbol{r}_2-\boldsymbol{r}'\right)\nonumber\\
 & +\int d\boldsymbol{r}_1d\boldsymbol{r}_2G_{AB}^{K'}\left(\boldsymbol{r}-\boldsymbol{r}_1\right)V_{BA}^{K'K}\left(\boldsymbol{r}_1,\boldsymbol{r}_2\right)G_{AA}^{K}\left(\boldsymbol{r}_2-\boldsymbol{r}'\right)\nonumber\\
 \delta G_{BB}^{KK'}\left(\boldsymbol{r},\boldsymbol{r}'\right)  & =\int d\boldsymbol{r}_1d\boldsymbol{r}_2G_{BB}^{K}\left(\boldsymbol{r}-\boldsymbol{r}_1\right)V_{BA}^{KK'}\left(\boldsymbol{r}_1,\boldsymbol{r}_2\right)G_{AB}^{K'}\left(\boldsymbol{r}_2-\boldsymbol{r}'\right)\nonumber\\
 & +\int d\boldsymbol{r}_1d\boldsymbol{r}_2G_{BA}^{K}\left(\boldsymbol{r}-\boldsymbol{r}_1\right)V_{AB}^{KK'}\left(\boldsymbol{r}_1,\boldsymbol{r}_2\right)G_{BB}^{K'}\left(\boldsymbol{r}_2-\boldsymbol{r}'\right)\nonumber\\
\delta G_{BB}^{K'K}\left(\boldsymbol{r},\boldsymbol{r}'\right)  & =\int d\boldsymbol{r}_1d\boldsymbol{r}_2 G_{BB}^{K'}\left(\boldsymbol{r}-\boldsymbol{r}_1\right)V_{BA}^{K'K}\left(\boldsymbol{r}_1,\boldsymbol{r}_2\right)G_{AB}^{K}\left(\boldsymbol{r}_2-\boldsymbol{r}'\right)\nonumber\\
 & +\int d\boldsymbol{r}_1d\boldsymbol{r}_2G_{BA}^{K'}\left(\boldsymbol{r}-\boldsymbol{r}_1\right)V_{AB}^{K'K}\left(\boldsymbol{r}_1,\boldsymbol{r}_2\right)G_{BB}^{K}\left(\boldsymbol{r}_2-\boldsymbol{r}'\right).\nonumber
\end{align}
We use the Green's function matrix elements (\ref{eq:graphene Green function}) and the vacancy potential $\cal{V}$ matrix element of (8) in the main text with $\boldsymbol{R}_0=\boldsymbol{0}$, defined by the $4 \times 4$ matrix $V_A\left(\boldsymbol{r}_{1},\boldsymbol{r}_{2}\right) \equiv   \langle \boldsymbol{r}_{1} | {\cal V} | \boldsymbol{r}_{2} \rangle$ 
\begin{align}
V_{A}\left(\boldsymbol{r}_{1},\boldsymbol{r}_{2}\right) & = a^2  v_F\left(\begin{array}{cccc}
0 & 0 & 0 & \delta\left(\boldsymbol{r}_{1}\right)L_{\boldsymbol{r}_{2}}^{\dagger}\\
0 & 0 & -\delta\left(\boldsymbol{r}_{1}\right)L_{\boldsymbol{r}_{2}} & 0\\
0 & \delta\left(\boldsymbol{r}_{2}\right)L_{\boldsymbol{r}_{1}}^{\dagger} & 0 & 0\\
-\delta\left(\boldsymbol{r}_{2}\right)L_{\boldsymbol{r}_{1}} & 0 & 0 & 0
\end{array}\right)\delta\left(\boldsymbol{r}_{1}-\boldsymbol{r}_{2}\right) \, .
\end{align}
Let us calculate (\ref{eq:first order valley and sublattice contributions}) for graphene with a vacancy as given by Hamiltonian (\ref{eq:graphene with vacancy Hamitonian})

\begin{align}
\delta G_{AA}^{KK'}\left(\boldsymbol{r},\boldsymbol{r}'\right)  & =-a^2v_{F}\int d\boldsymbol{r}_{1}d\boldsymbol{r}_{2}G_{AA}^{K}\left(\boldsymbol{r}-\boldsymbol{r}_{1}\right)\delta\left(\boldsymbol{r}_{1}\right)\delta\left(\boldsymbol{r}_{1}-\boldsymbol{r}_{2}\right)\left[L_{\boldsymbol{r}_{2}}^{\dagger}G_{BA}^{K'}\left(\boldsymbol{r}_{2}-\boldsymbol{r}'\right)\right]\\
 & -a^2v_{F}\int d\boldsymbol{r}_{1}d\boldsymbol{r}_{2}\left[L_{\boldsymbol{r}_{1}}^{\dagger}G_{AB}^{K}\left(\boldsymbol{r}-\boldsymbol{r}_{1}\right)\right]\delta\left(\boldsymbol{r}_{2}\right)\delta\left(\boldsymbol{r}_{1}-\boldsymbol{r}_{2}\right)G_{AA}^{K'}\left(\boldsymbol{r}_{2}-\boldsymbol{r}'\right)\nonumber \\
\delta G_{AA}^{K'K}\left(\boldsymbol{r},\boldsymbol{r}'\right) & =a^2v_{F}\int d\boldsymbol{r}_{1}d\boldsymbol{r}_{2}G_{AA}^{K'}\left(\boldsymbol{r}-\boldsymbol{r}_{1}\right)\delta\left(\boldsymbol{r}_{1}\right)\delta\left(\boldsymbol{r}_{1}-\boldsymbol{r}_{2}\right)\left[L_{\boldsymbol{r}_{2}}G_{BA}^{K}\left(\boldsymbol{r}_{2}-\boldsymbol{r}'\right)\right]\nonumber \\
 & +a^2v_{F}\int d\boldsymbol{r}_{1}d\boldsymbol{r}_{2}\left[L_{\boldsymbol{r}_{1}}G_{AB}^{K'}\left(\boldsymbol{r}-\boldsymbol{r}_{1}\right)\right]\delta\left(\boldsymbol{r}_{2}\right)\delta\left(\boldsymbol{r}_{1}-\boldsymbol{r}_{2}\right)G_{AA}^{K}\left(\boldsymbol{r}_{2}-\boldsymbol{r}'\right)\nonumber \\
\delta G_{BB}^{KK'}\left(\boldsymbol{r},\boldsymbol{r}'\right) & =-a^2v_{F}\int d\boldsymbol{r}_{1}d\boldsymbol{r}_{2}\left[L_{\boldsymbol{r}_{1}}^{\dagger}G_{BB}^{K}\left(\boldsymbol{r}-\boldsymbol{r}_{1}\right)\right]\delta\left(\boldsymbol{r}_{2}\right)\delta\left(\boldsymbol{r}_{1}-\boldsymbol{r}_{2}\right)G_{AB}^{K'}\left(\boldsymbol{r}_{2}-\boldsymbol{r}'\right)\nonumber \\
 & -a^2v_{F}\int d\boldsymbol{r}_{1}d\boldsymbol{r}_{2}G_{BA}^{K}\left(\boldsymbol{r}-\boldsymbol{r}_{1}\right)\delta\left(\boldsymbol{r}_{1}\right)\delta\left(\boldsymbol{r}_{1}-\boldsymbol{r}_{2}\right)\left[L_{\boldsymbol{r}_{2}}^{\dagger}G_{BB}^{K'}\left(\boldsymbol{r}_{2}-\boldsymbol{r}'\right)\right]\nonumber \\
\delta G_{BB}^{K'K}\left(\boldsymbol{r},\boldsymbol{r}'\right) & =a^2v_{F}\int d\boldsymbol{r}_{1}d\boldsymbol{r}_{2}\left[L_{\boldsymbol{r}_{1}}G_{BB}^{K'}\left(\boldsymbol{r}-\boldsymbol{r}_{1}\right)\right]\delta\left(\boldsymbol{r}_{2}\right)\delta\left(\boldsymbol{r}_{1}-\boldsymbol{r}_{2}\right)G_{AB}^{K}\left(\boldsymbol{r}_{2}-\boldsymbol{r}'\right)\nonumber \\
 & +a^2v_{F}\int d\boldsymbol{r}_{1}d\boldsymbol{r}_{2}G_{BA}^{K'}\left(\boldsymbol{r}-\boldsymbol{r}_{1}\right)\delta\left(\boldsymbol{r}_{1}\right)\delta\left(\boldsymbol{r}_{1}-\boldsymbol{r}_{2}\right)\left[L_{\boldsymbol{r}_{2}}G_{BB}^{K}\left(\boldsymbol{r}_{2}-\boldsymbol{r}'\right)\right]\nonumber 
\end{align}
where we integrated by parts. Using  $L_{\boldsymbol{r}_{2}}^{\dagger}G_{BA}^{K'}\left(\boldsymbol{r}_{2}-\boldsymbol{r}'\right)=-L_{\boldsymbol{r}'}^{\dagger}G_{BA}^{K'}\left(\boldsymbol{r}_{2}-\boldsymbol{r}'\right)$
and performing integration over the two delta functions leads to,
\begin{align}
\delta G_{AA}^{KK'}\left(\boldsymbol{r},\boldsymbol{r}'\right) & =a^2v_{F}\left(G_{AA}^{K}\left(\boldsymbol{r}\right)\left(L_{\boldsymbol{r}'}^{\dagger}G_{BA}^{K'}\left(-\boldsymbol{r}'\right)\right) +\left(L_{\boldsymbol{r}}^{\dagger}G_{AB}^{K}\left(\boldsymbol{r}\right)\right)G_{AA}^{K'}\left(-\boldsymbol{r}'\right)\right)\\
\delta G_{AA}^{K'K}\left(\boldsymbol{r},\boldsymbol{r}'\right)  & =-a^2v_{F}\left(G_{AA}^{K'}\left(\boldsymbol{r}\right)\left(L_{\boldsymbol{r}'}G_{BA}^{K}\left(-\boldsymbol{r}'\right)\right)\nonumber  +\left(L_{\boldsymbol{r}}G_{AB}^{K'}\left(\boldsymbol{r}\right)\right)G_{AA}^{K}\left(-\boldsymbol{r}'\right)\right)\nonumber \\
\delta G_{BB}^{KK'}\left(\boldsymbol{r},\boldsymbol{r}'\right)  & =a^2v_{F}\left(\left(L_{\boldsymbol{r}}^{\dagger}G_{BB}^{K}\left(\boldsymbol{r}\right)\right)G_{AB}^{K'}\left(-\boldsymbol{r}'\right)\nonumber +G_{BA}^{K}\left(\boldsymbol{r}\right)\left(L_{\boldsymbol{r}'}^{\dagger}G_{BB}^{K'}\left(-\boldsymbol{r}'\right)\right)\right)\nonumber \\
\delta G_{BB}^{K'K}\left(\boldsymbol{r},\boldsymbol{r}'\right) & =-a^2v_{F}\left(\left(L_{\boldsymbol{r}}G_{BB}^{K'}\left(\boldsymbol{r}\right)\right)G_{AB}^{K}\left(-\boldsymbol{r}'\right)\nonumber 
  +G_{BA}^{K'}\left(\boldsymbol{r}\right)\left(L_{\boldsymbol{r}'}G_{BB}^{K}\left(-\boldsymbol{r}'\right)\right)\right)\nonumber \, .
\end{align}
These equations simplify by inserting Green's functions of two-valleys pristine graphene \eqref{eq:graphene Green function}, and using that $\tilde{g}_{\epsilon}\left(r\right)$
is symmetric with $r$, so that $L_{\boldsymbol{r}}^{\dagger}\tilde{g}_{\epsilon}\left(-r\right)=-L_{\boldsymbol{r}}^{\dagger}\tilde{g}_{\epsilon}\left(r\right)$. This calculation simplifies by performing the sum over the two sublattices terms. We are specifically interested in the case of $\boldsymbol{r}'= \boldsymbol{r}$ so that,
\begin{align}
\delta G^{KK'}_{\epsilon}\left(\boldsymbol{r}\right)  & =2a^2\epsilon \,  e^{i\Delta\boldsymbol{K}\cdot\boldsymbol{r}}\left[\tilde{g}_{\epsilon}\left(r\right)\left(L_{\boldsymbol{r}}^{\dagger}L_{\boldsymbol{r}}^{\vphantom{\ast}}\tilde{g}_{\epsilon}\left(r\right)\right)-\left(L_{\boldsymbol{r}}^{\dagger}\tilde{g}_{\epsilon}\left(r\right)\right)^{2}\right]\\
\delta G^{K'K}_{\epsilon}\left(\boldsymbol{r}\right)  & =2a^2\epsilon \,  e^{-i\Delta\boldsymbol{K}\cdot\boldsymbol{r}}\left[\tilde{g}_{\epsilon}\left(r\right)\left(L_{\boldsymbol{r}}^{\vphantom{\ast}}L_{\boldsymbol{r}}^{\dagger}\frac{\epsilon}{v_{F}}\tilde{g}_{\epsilon}\left(r\right)\right)-\left(L_{\boldsymbol{r}}\tilde{g}_{\epsilon}\left(r\right)\right)^{2}\right]\nonumber 
\end{align}
where for graphene $\boldsymbol{K}=-\boldsymbol{K}'$. Using the relations, 
\begin{equation}
\begin{cases}
L_{\boldsymbol{r}}^{\dagger}L_{\boldsymbol{r}}^{\vphantom{\ast}}=L_{\boldsymbol{r}}^{\vphantom{\ast}}L_{\boldsymbol{r}}^{\dagger}=-\nabla^{2}\\
L_{\boldsymbol{r}}\tilde{g}_{\epsilon}\left(r\right)=-ie^{-i\theta}\frac{d}{dr}\tilde{g}_{\epsilon}\left(r\right)\\
L_{\boldsymbol{r}}^{\dagger}\tilde{g}_{\epsilon}\left(r\right)=-ie^{i\theta}\frac{d}{dr}\tilde{g}_{\epsilon}\left(r\right)
\end{cases}
\end{equation}
we obtain expression (14) of the main text. Finally, summing over the two valleys and integrating over energies, we  obtain the local electronic density $\delta\rho_A \left(\boldsymbol{r}\right)= -\frac{1}{\pi} \int d\epsilon \, \text{Im} \, \delta G_\epsilon ^{R}\left(\boldsymbol{r},\boldsymbol{r} \right)   
$. Dividing this expression into two integrals 
\begin{equation}
a^{2}\, \text{Im}\int_{-\epsilon_{min}}^{0}d\epsilon \, \epsilon \, \tilde{g}_{\epsilon}\left(r\right)\nabla^{2}\tilde{g}_{\epsilon}\left(r\right)=-\frac{a^{2}}{16}\text{Im}\int_{-\epsilon_{min}}^{0}d\epsilon \, \frac{\epsilon}{v_{F}^{2}}H_{0}^{\left(1\right)}\left(\frac{\epsilon r}{v_{F}}\right)\nabla^{2}H_{0}^{\left(1\right)}\left(\frac{\epsilon r}{v_{F}}\right)
\end{equation}
\begin{equation}
a^{2} \, \text{Im}\int d\epsilon \, \epsilon \, \left(\frac{d}{dr}\tilde{g}_{\epsilon}\left(r\right)\right)^{2}=-\frac{a^{2}}{16}\text{Im}\int d\epsilon\frac{\epsilon}{v_{F}^{2}}\left(\frac{d}{dr}H_{0}^{\left(1\right)}\left(\frac{\epsilon r}{v_{F}}\right)\right)^{2}
\end{equation}
where $\epsilon_{min}=3t$. The change of variables $z=\frac{\epsilon r}{v_{F}}$, leads to 
\begin{align}
a^{2}\, \text{Im}\int_{-\epsilon_{min}}^{0}d\epsilon \, \epsilon \, \tilde{g}_{\epsilon}\left(r\right)\nabla^{2}\tilde{g}_{\epsilon}\left(r\right) & =-\frac{a^{2}}{16r^{4}}\text{Im}\int_{-2r/a}^{0}dzz^{3}H_{0}^{\left(1\right)}\left(z\right)\nabla_{z}^{2}H_{0}^{\left(1\right)}\left(z\right)
\end{align}
\begin{align}
a^{2}\, \text{Im}\int d\epsilon \, \epsilon\left(\frac{d}{dr}\tilde{g}_{\epsilon}\left(r\right)\right)^{2} & =-\frac{a^{2}}{16 r^{4}}\text{Im}\int_{-2r/a}^{0}dzz^{3}\left(H_{1}^{\left(1\right)}\left(z\right)\right)^{2}.
\end{align}
Another change of variables, $u=z\frac{a}{r}$, removes the $r/a$ dependence from the integral boundaries. For $r/a \gg 1$, the upper limit does not contribute, hence we can use the asymptotic limit of the Hankel functions $H_\nu^{(1)}\left(x\right)\simeq\sqrt{\frac{2}{\pi x}}e^{i(x-\nu\frac{\pi}{2}-\frac{\pi}{4})}$,   
\begin{align}
a^{2}\text{Im}\int_{-\epsilon_{min}}^{0}d\epsilon \, \epsilon \, \tilde{g}_{\epsilon}\left(r\right)\nabla^{2}\tilde{g}_{\epsilon}\left(r\right) & \simeq-\frac{1}{4\pi r^{2}}\sin\left(4r/a\right)
\end{align}
\begin{equation}
a^{2}\, \text{Im}\int d\epsilon \, \epsilon\left(\frac{d}{dr}\tilde{g}_{\epsilon}\left(r\right)\right)^{2} \simeq- \frac{1}{4\pi{r^{2}}}\sin\left(4r/a\right) \, .
\end{equation}
This leads to expression (1) of the main text for the local density with two oscillating planar waves terms, 
\begin{align}\label{eq:local density vacancy}
\delta\rho_{A}^{(1)}\left(\boldsymbol{r}\right) & =\frac{1}{\pi^{2}r^{2}}\sin\left(4r/a\right)\left[\cos\left(\Delta\boldsymbol{K}\cdot\boldsymbol{r}+2\theta\right)-\cos\Delta\boldsymbol{K}\cdot\boldsymbol{r}\right] \, .
\end{align}

\section{Higher order scattering}
In this section, we establish expression (1) in the main text for all order in perturbation theory, namely they are all proportional to the first order up to a radial function. We discuss only odd orders since they contribute to inter-valley scattering. This is a property of the potential $\cal V$  which contains only $K\to K'$ terms. For graphene with local defect (proportional to a delta function), propagators of the type  $G_{AB/BA}^{K/K'}$ vanish when inserted between two $\cal V$ operators. This is due to $G_{AB/BA}^{K/K'}$ being an anti-symmetric function of $x,y$  meaning  $G_{AB/BA}^{K/K'}$ at $r=0$ vanishes. Hence all scattering processes can be divided into contributions of the first-order term multiplied by a repeating series of operators. For example, the generic   $AA$ scattering term is
\begin{align}\label{eq:Green function block form}
\delta G_{AA}^{KK'\left(2n+1\right)} & =\left(G_{AA}^{K}V_{AB}^{KK'}G_{BB}^{K'K'}V_{BA}^{K'K}\right)^{n}G_{AA}^{K}V_{AB}^{KK'}G_{BA}^{K'}\\
 & +G_{AB}^{K}V_{BA}^{KK'}G_{AA}^{K'}\left(V_{AB}^{K'K}G_{BB}^{K}V_{BA}^{KK'}G_{AA}^{K'}\right)^{n}\nonumber 
\end{align}
The repeating terms, e.g.  $G_{AA}^{K}V_{AB}^{KK'}G_{BB}^{K'K'}V_{BA}^{K'K}$, are always composed of products of four terms : two diagonal elements of the sublattice $G_{AA/BB}$ which do not contain $L_{\boldsymbol{r}}$ or $L^{\dagger}_{\boldsymbol{r}}$ and another two complex conjugate operators, such that for each $L_{\boldsymbol{r}}$ operator there is an $L^{\dagger}_{\boldsymbol{r}}$ operator. These pairs of $L_{\boldsymbol{r}}^{\vphantom{\ast}}$, $L^{\dagger}_{\boldsymbol{r}}$ operators cancel the $\theta$ dependence of the repeating blocks. Finally, we obtain the matrix elements,
\begin{align}
\left\langle \boldsymbol{r}\right|G_{AA}^{K}V_{AB}^{KK'}G_{BB}^{K'K'}V_{BA}^{K'K}\left|\boldsymbol{r}'\right\rangle = & \left(a^{2}v_{F}\right)^{2}G_{AA}^{K}\left(\boldsymbol{r}\right)\left[L_{\boldsymbol{r}'}^{\dagger}L_{\boldsymbol{r}'}^{\vphantom{\ast}}G_{BB}^{K'}\left(\boldsymbol{r}'\right)\right]\delta\left(\boldsymbol{r}'\right)\\
\left\langle \boldsymbol{r}\right|V_{AB}^{K'K}G_{BB}^{K}V_{BA}^{KK'}G_{AA}^{K'}\left|\boldsymbol{r}'\right\rangle = & \left(a^{2}v_{F}\right)^{2}\delta\left(\boldsymbol{r}\right)\left[L_{\boldsymbol{r}}^{\dagger}L_{\boldsymbol{r}}^{\vphantom{\ast}}G_{BB}^{K}\left(\boldsymbol{r}\right)\right]G_{AA}^{K'}\left(-\boldsymbol{r}'\right)\nonumber 
\end{align}
and their relation to the $n^{\text th}$ term 
\begin{align}\label{'eq:higer order blocks'}
\left\langle \boldsymbol{r}\right|\left(G_{AA}^{K}V_{AB}^{KK'}G_{BB}^{K'K'}V_{BA}^{K'K}\right)^{n}\left|\boldsymbol{r}'\right\rangle  & =\left(C^{K'}\right)^{n-1}\left\langle \boldsymbol{r}\right|G_{AA}^{K}V_{AB}^{KK'}G_{BB}^{K'K'}V_{BA}^{K'K}\left|\boldsymbol{r}'\right\rangle \\
\left\langle \boldsymbol{r}\right|\left(V_{AB}^{K'K}G_{BB}^{K}V_{BA}^{KK'}G_{AA}^{K'}\right)^{n}\left|\boldsymbol{r}'\right\rangle  & =\left(C^{K}\right)^{n-1}\left\langle \boldsymbol{r}\right|V_{AB}^{K'K}G_{BB}^{K}V_{BA}^{KK'}G_{AA}^{K'}\left|\boldsymbol{r}'\right\rangle \nonumber 
\end{align}
where we define two dimensionless scaling functions, 
\begin{equation}
    C^{K'/K}\left(\frac{a\epsilon}{v_F} \right
)\equiv\left(a^{2}v_{F}\right)^2\left[\left.L_{\boldsymbol{r}_{1}}^{\dagger}L_{\boldsymbol{r}_{1}}^{\vphantom{\ast}}G_{BB}^{K'/K}\left(\boldsymbol{r}_{1}\right)\right|_{\boldsymbol{r}_{1}=0}\right]G_{AA}^{K,K'}\left(0\right).
\end{equation}
 These two functions are equal for   $\left|\boldsymbol{K}\right|=\left|\boldsymbol{K}'\right|$, since the first derivative  $L_{\boldsymbol{r}_1}G_{BB}^{K/K'}$ vanishes at the origin. We examine the $(2n+1)$-contribution to the Green's function. We use \eqref{'eq:higer order blocks'} to calculate the matrix elements of \eqref{eq:Green function block form} 
\begin{align}
\left\langle \boldsymbol{r}\right|\delta G_{AA}^{KK'\left(2n+1\right)}\left|\boldsymbol{r}'\right\rangle  & =\left(C^{K}\right)^{n-1}\int d\boldsymbol{r}_{1}\left\langle \boldsymbol{r}\right|G_{AA}^{K}V_{AB}^{KK'}G_{BB}^{K'K'}V_{BA}^{K'K}\left|\boldsymbol{r}_{1}\right\rangle \left\langle \boldsymbol{r}_{1}\right|G_{AA}^{K}V_{AB}^{KK'}G_{BA}^{K'}\left|\boldsymbol{r}'\right\rangle \\
 & +\left(C^{K}\right)^{n-1}\int d\boldsymbol{r}_{1}\left\langle \boldsymbol{r}\right|G_{AB}^{K}V_{BA}^{KK'}G_{AA}^{K'}\left|\boldsymbol{r}_{1}\right\rangle \left\langle \boldsymbol{r}_{1}\right|V_{AB}^{K'K}G_{BB}^{K}V_{BA}^{KK'}G_{AA}^{K'}\left|\boldsymbol{r}'\right\rangle \nonumber \, .
\end{align}
Integrating over $\boldsymbol{r}_1$ 
and inserting all the matrix elements previously calculated provide an identical contribution as the first order with the exception of the $\left(C^{K}\right)^{n}$ term,
\begin{align}
\left\langle \boldsymbol{r}\right|\delta G_{AA}^{KK'\left(2n+1\right)}\left|\boldsymbol{r}'\right\rangle  & =\left(C^{K}\right)^{n}a^{2}\epsilon \,  e^{i\boldsymbol{K}\cdot\boldsymbol{r}}\tilde{g}_{\epsilon}\left(r\right)\left(L_{\boldsymbol{r}'}^{\dagger}e^{-i\boldsymbol{K}'\cdot\boldsymbol{r}'}L_{\boldsymbol{r}'}^{\vphantom{\ast}}\tilde{g}_{\epsilon}\left(-r'\right)\right)\\
 & +\left(C^{K}\right)^{n}a^{2}\epsilon \, e^{-i\boldsymbol{K}'\cdot\boldsymbol{r}'}\left(L_{\boldsymbol{r}}^{\dagger}e^{i\boldsymbol{K}\cdot\boldsymbol{r}}L_{\boldsymbol{r}}^{\vphantom{\ast}}\tilde{g}_{\epsilon}\left(r\right)\right)\tilde{g}_{\epsilon}\left(-r'\right).\nonumber 
\end{align}\\

This calculation repeats itself for all vacancy scattering processes. Integrating over energy provides two identical integrals to learning order in $a/r$,
\begin{align}
\int_{\epsilon_{min}}^{0}d\epsilon\left(C^{K}\left(\frac{a\epsilon}{v_{F}}\right)\right)^{n}a^{2}\epsilon\left(\tilde{g}_{\epsilon}\left(r\right)\nabla_{r}^{2}\tilde{g}_{\epsilon}\left(r\right)\right) & =\frac{1}{8\pi r}\frac{1}{a}\int_{-2}^{0}du\left(C^{K}\left(u\right)\right)^{n}e^{2i(\frac{r}{a}u-\frac{\pi}{4})}u^{2}\\
\int_{\epsilon_{min}}^{0}d\epsilon\left(C^{K}\left(\frac{a\epsilon}{v_{F}}\right)\right)^{n}a^{2}\epsilon\left(\left(\frac{d}{dr}\tilde{g}_{\epsilon}\left(r\right)\right)^{2}\right) & =\frac{1}{8\pi r}\frac{1}{a}\int_{-2}^{0}du\left(C^{K}\left(u\right)\right)^{n}e^{2i(\frac{r}{a}u-\frac{\pi}{4})}u^{2}\nonumber 
\end{align}
meaning the two radial function weighting $\cos\left(\Delta \boldsymbol{K}\cdot \boldsymbol{r}\right)$ and  $\cos\left(\Delta \boldsymbol{K}\cdot \boldsymbol{r}+2\theta\right)$ are equal, hence leading to an interference pattern to all orders, thus generalising the first  order expression, namely 
\begin{align}
\delta\rho\left(\boldsymbol{r}\right) & =F\left(r\right)\left[\cos\left(\Delta\boldsymbol{K}\cdot\boldsymbol{r}+2\theta\right)-\cos\Delta\boldsymbol{K}\cdot\boldsymbol{r}\right].
\end{align}
where $F\left(r\right)=\sum_{n=0}^{\infty}\frac{1}{8\pi r}\frac{1}{a}\int_{-2}^{0}du\left(C^{K}\left(u\right)\right)^{n}e^{2i(\frac{r}{a}u-\frac{\pi}{4})}u^{2}$.

\section{Topological equivalence between a vacancy and the Kekule model} 
In this section, we establish the topological equivalence between graphene with a vacancy and the Kekule distortion defined in the main text. 

In the continuum limit \cite{Hou2007}, the Kekule distortion model can be represented by an effective massless fermion field $\psi_{\boldsymbol{r}}$, coupled to a complex scalar field such that the axial symmetry is preserved, with the Hamiltonian 
\begin{align}\label{eq:HCM_hamiltonian}
H=\int d\boldsymbol{r}\, \psi^{\dagger}_{\boldsymbol{r}}{\cal{K}} \psi_{\boldsymbol{r}},
\end{align}
\begin{equation}
\cal{K}=\left(\begin{array}{cccc}
0 & v_F\left(-i\partial_{x}-\partial_{y}\right) & \Delta\left(\boldsymbol{r}\right) & 0\\
v_F\left(-i\partial_{x}+\partial_{y}\right) & 0 & 0 & \Delta\left(\boldsymbol{r}\right)\\
\Delta^{*}\left(\boldsymbol{r}\right) & 0 & 0 & -v_F\left(-i\partial_{x}-\partial_{y}\right)\\
0 & \Delta^{*}\left(\boldsymbol{r}\right) & -v_F\left(-i\partial_{x}+\partial_{y}\right) & 0\nonumber
\end{array}\right)
\end{equation}
in the basis $\left(\begin{array}{cccc}
\psi_{A}^{+} & \psi_{B}^{+} & \psi_{A}^{-} & \psi_{B}^{-}\end{array}\right)^{T}$. 
 To relate ${\cal{K}}$ to $H_0 + {\cal V}$, we perform a unitary transformation, transforming the basis to $\left(\begin{array}{cccc}
\psi_{A}^{-} & \psi_{A}^{+} & -\psi_{B}^{-} & -\psi_{B}^{+}\end{array}\right)$. In this basis, the contributions of $H_0$ are similar as in graphene with a vacancy,
\begin{equation}\label{eq:Kekule Hamiltonian}
\tilde{\cal{K}}=\left(\begin{array}{cccc}
0 & 0 & v_FL_{\boldsymbol{r}} & -\Delta_{0}\left(r\right)e^{-in\theta-i\alpha}\\
0 & 0 & -\Delta_{0}^{*}\left(r\right)e^{in\theta+i\alpha} & -v_FL_{\boldsymbol{r}}^{\dagger}\\
v_FL_{\boldsymbol{r}}^{\dagger} & -\Delta_{0}\left(r\right)e^{-in\theta-i\alpha} & 0 & 0\\
-\Delta_{0}^{*}\left(r\right)e^{in\theta+i\alpha} & -v_FL_{\boldsymbol{r}} & 0 & 0
\end{array}\right).
\end{equation}
Defining $\tilde{\cal{K}} \equiv H_0 + {\cal V}_K$, we thus recover that ${\cal V}_K$ and $\cal V$ have the same structure as announced in the text.

\section{Local density for the Kekule Hamiltonian}
We present a proof of (3) for a Kekule distortion. To that purpose, we  calculate (\ref{eq:first order valley and sublattice contributions}) for the Kekule Hamiltonian (\ref{eq:Kekule Hamiltonian}). We define \begin{equation}
h\left(\boldsymbol{r},\boldsymbol{r}_1\right)\equiv e^{i\Delta\boldsymbol{K}\cdot \left(\boldsymbol{r}-\boldsymbol{r}_1\right)-i\alpha} \Delta_0\left(r_1\right)e^{-in\theta_1}    
\end{equation} such that summing over the two scattering processes $KK'+K'K$ for each of the two sublattices provides 
\begin{align}
\delta G_{AA}^{\left(1\right)}\left(\boldsymbol{r}\right)  & =2\frac{\epsilon}{v_{F}}\int d\boldsymbol{r}_{1}h\left(\boldsymbol{r},\boldsymbol{r}_{1}\right)\left(\tilde{g}_{\epsilon}\left(\left|\boldsymbol{r}_{1}-\boldsymbol{r}\right|\right)\left(-i\frac{d}{dx_{1}}-\frac{d}{dy_{1}}\right)\tilde{g}_{\epsilon}\left(\left|\boldsymbol{r}_{1}-\boldsymbol{r}\right|\right)\right)\\
 & +2\frac{\epsilon}{v_{F}}\int d\boldsymbol{r}_{1}h^{*}\left(\boldsymbol{r},\boldsymbol{r}_{1}\right)\left(\tilde{g}_{\epsilon}\left(\left|\boldsymbol{r}_{1}-\boldsymbol{r}\right|\right)\left(i\frac{d}{dx_{1}}-\frac{d}{dy_{1}}\right)\tilde{g}_{\epsilon}\left(\left|\boldsymbol{r}_{1}-\boldsymbol{r}\right|\right)\right)\nonumber 
\end{align}
\begin{align}
\delta G_{BB}^{\left(1\right)}\left(\boldsymbol{r}\right) & =2\frac{\epsilon}{v_{F}}\int d\boldsymbol{r}_{1}h\left(\boldsymbol{r},\boldsymbol{r}_{1}\right)\left(\tilde{g}_{\epsilon}\left(\left|\boldsymbol{r}_{1}-\boldsymbol{r}\right|\right)\left(-i\frac{d}{dx_{1}}+\frac{d}{dy_{1}}\right)\tilde{g}_{\epsilon}\left(\left|\boldsymbol{r}_{1}-\boldsymbol{r}\right|\right)\right)\\
 & +2\frac{\epsilon}{v_{F}}\int d\boldsymbol{r}_{1}h^{*}\left(\boldsymbol{r},\boldsymbol{r}_{1}\right)\left(\tilde{g}_{\epsilon}\left(\left|\boldsymbol{r}_{1}-\boldsymbol{r}\right|\right)\left(i\frac{d}{dx_{1}}+\frac{d}{dy_{1}}\right)\tilde{g}_{\epsilon}\left(\left|\boldsymbol{r}_{1}-\boldsymbol{r}\right|\right)\right)\nonumber \, .
\end{align}
By taking the derivative with respect to $x_1$ or $y_1$, we further simplify the two expression using their definition $\tilde{g}_{\epsilon}\left(\left|\boldsymbol{r}_{1}-\boldsymbol{r}\right|\right)=-i\frac{1}{v_{F}}H_{0}^{\left(1\right)}\left(\frac{\epsilon\left|\boldsymbol{r}_{1}-\boldsymbol{r}\right|}{v_{F}}\right)$
, $\tilde{g}'_{\epsilon}\left(\left|\boldsymbol{r}_{1}-\boldsymbol{r}\right|\right)=i\frac{1}{v_{F}}H_{1}^{\left(1\right)}\left(\frac{\epsilon\left|\boldsymbol{r}_{1}-\boldsymbol{r}\right|}{v_{F}}\right)$, namely,
\begin{align}
\delta G_{AA}^{\left(1\right)}\left(\boldsymbol{r}\right)  & =4\int d\boldsymbol{r}_{1}\left(\text{Im}\, h\left(\boldsymbol{r},\boldsymbol{r}_{1}\right)\frac{x_{1}-x}{\left|\boldsymbol{r}_{1}-\boldsymbol{r}\right|}-\text{Re}\, h\left(\boldsymbol{r},\boldsymbol{r}_{1}\right)\frac{y_{1}-y}{\left|\boldsymbol{r}_{1}-\boldsymbol{r}\right|}\right)\frac{\epsilon^2}{v_F^2}\tilde{g}_{\epsilon}\left(\left|\boldsymbol{r}_{1}-\boldsymbol{r}\right|\right)\tilde{g}'_{\epsilon}\left(\left|\boldsymbol{r}_{1}-\boldsymbol{r}\right|\right)\\
\delta G_{BB}\left(\boldsymbol{r}\right) & =4\int d\boldsymbol{r}_{1}\left(\text{Im}\, h\left(\boldsymbol{r},\boldsymbol{r}_{1}\right)\frac{x_{1}-x}{\left|\boldsymbol{r}_{1}-\boldsymbol{r}\right|}+\text{Re}\, h\left(\boldsymbol{r},\boldsymbol{r}_{1}\right)\frac{y_{1}-y}{\left|\boldsymbol{r}_{1}-\boldsymbol{r}\right|}\right)\frac{\epsilon^2}{v_F^2}\tilde{g}_{\epsilon}\left(\left|\boldsymbol{r}_{1}-\boldsymbol{r}\right|\right)\tilde{g}'_{\epsilon}\left(\left|\boldsymbol{r}_{1}-\boldsymbol{r}\right|\right).\nonumber
\end{align}
The real and imaginary parts of $h$ are 
\begin{align}
\text{Re}\, h\left(\boldsymbol{r},\boldsymbol{r}_{1}\right) & =\Delta_0\left(r_1\right)\cos\left(\Delta \boldsymbol{K}\cdot\left(\boldsymbol{r}-\boldsymbol{r}_1\right)-n\theta_1-\alpha\right)\\
\text{Im}\, h\left(\boldsymbol{r},\boldsymbol{r}_{1}\right) & =\Delta_0\left(r_1\right)\sin\left(\Delta \boldsymbol{K}\cdot\left(\boldsymbol{r}-\boldsymbol{r}_1\right)-n\theta_1-\alpha\right)\nonumber \, .
\end{align}
In addition, we define $\boldsymbol{R}\equiv\boldsymbol{r}_{1}-\boldsymbol{r}$,
so that $\frac{x_{1}-x}{\left|\boldsymbol{r}_{1}-\boldsymbol{r}\right|} \equiv \cos\theta_{R}$ 
and $\frac{y_{1}-y}{\left|\boldsymbol{r}_{1}-\boldsymbol{r}\right|} \equiv \sin\theta_{R}$. These relations help us simplify the two sublattices scattering process using trigonometric identities. Integrating over  energies, we obtain

\begin{align} \label{hard}
\delta \rho_{AA}\left(\boldsymbol{r}\right) & =-\frac{4}{\pi}\int d\boldsymbol{r}_{1}\Delta_{0}\left(r_{1}\right)\sin\left(\Delta\boldsymbol{K}\cdot\boldsymbol{R}-n\theta_{1}-\theta_{R}-\alpha\right)\int d\epsilon\text{Im}\frac{\epsilon^2}{v_F^2}\tilde{g}_{\epsilon}\left({R}\right)\tilde{g}'_{\epsilon}\left(R\right)\\
\delta \rho_{BB}\left(\boldsymbol{r}\right) & =-\frac{4}{\pi}\int d\boldsymbol{r}_{1}\Delta_{0}\left(r_{1}\right)\sin\left(\Delta\boldsymbol{K}\cdot\boldsymbol{R}-n\theta_{1}+\theta_{R}-\alpha\right)\int d\epsilon\text{Im}\frac{\epsilon^2}{v_F^2}\tilde{g}_{\epsilon}\left({R}\right)\tilde{g}'_{\epsilon}\left(R\right)\nonumber \, .
\end{align}
%The two integrals we obtained are not trivial since if one chose the two angles $\theta_1$ and $\theta_R$ (of $\boldsymbol{r}_1$ and $\boldsymbol{R}$ respectfully) depend on each other. In choosing one coordinate system, the other angle must be expressed regarding the other system's angle and radius. For this reason, we approach this problem numerically. However, 
We first examine the integral over energies, 
\begin{equation}
\int d\epsilon\text{Im}\frac{\epsilon^2}{v_F^2}\tilde{g}_{\epsilon}\left(R\right)\tilde{g}'_{\epsilon}\left(R\right)=\frac{1}{16}\text{Im}\int d\epsilon\frac{\epsilon^{2}}{v_{F}^{4}}H_{0}^{\left(1\right)}\left(\frac{\epsilon R}{v_{F}}\right)H_{1}^{\left(1\right)}\left(\frac{\epsilon R}{v_{F}}\right)
\end{equation}
defining $z \equiv \frac{\epsilon R}{v_{F}}$
\begin{equation}
\int d\epsilon\text{Im}\frac{\epsilon^2}{v_F^2}\tilde{g}_{\epsilon}\left(R\right)\tilde{g}'_{\epsilon}\left(R\right)=-\frac{1}{32v_{F}}\frac{1}{R^{3}}\text{Im}\int_{-2R/a}^{0}dzz^{2}\frac{d}{dz}\left(H_{0}^{\left(1\right)}\left(z\right)\right)^{2} \, .
\end{equation}
Integrating by parts leads to
\begin{equation}
\int d\epsilon\text{Im}\frac{\epsilon^2}{v_F^2}\tilde{g}_{\epsilon}\left(R\right)\tilde{g}'_{\epsilon}\left(R\right)=\frac{1}{16v_{F}}\text{Im}\left(\frac{2}{a^{2}R}\left(H_{0}^{\left(1\right)}\left(-\frac{2R}{a}\right)\right)^{2}+\frac{1}{R^{3}}\int_{-2R/a}^{0}dzz\left(H_{0}^{\left(1\right)}\left(z\right)\right)^{2}\right) \, .
\end{equation}
The first term decays as $R^{-2}$ and vanishes at the origin, while
the second term decays as $R^{-3}$ since its integral contribution
does not decay, thus to leading order we obtain
\begin{equation}
\int d\epsilon\text{Im}\frac{\epsilon^2}{v_F^2}\tilde{g}_{\epsilon}\left(R\right)\tilde{g}'_{\epsilon}\left(R\right)\simeq-\frac{1}{8\pi av_{F}}\frac{1}{R^{2}}\cos\left(\frac{4R}{a}\right)\equiv \tilde{F}\left(R\right) \, .
\end{equation}
This expression is valid for $R/a\gg1$. 
%onlysuch that at the origin there is no divergence. 

To perform the remaining integral in (\ref{hard}), we rotate $\boldsymbol{r}_1$ by an angle $\theta$  into $\tilde{\boldsymbol{r}}_1$, such that $\theta$ is still the angle between $\boldsymbol{r}$ and $\Delta\boldsymbol{K}$, but $\tilde{\theta}_1$ is the angle between $\boldsymbol{r}$ and $\boldsymbol{r}_1$. This means that $\boldsymbol{r}$ is aligned with $\boldsymbol{\tilde{x}}_1$, and $\theta_1=\tilde{\theta}_1+\theta$, $\theta_R=\tilde{\theta}_R+\theta$ where $\cos\tilde{\theta}_R=\frac{\tilde{x}_{1}-r}{\left|\boldsymbol{r}_{1}-\boldsymbol{r}\right|}$. In this coordinate system our integrals are 
\begin{align}
\delta\rho_{AA} ^{(1)} \left(\boldsymbol{r}\right) & =-\frac{4}{\pi}\int d\tilde{\boldsymbol{r}}_{1}\Delta_{0}\left(\tilde{r}_{1}\right)\sin\left(\Delta\boldsymbol{K}\cdot\boldsymbol{r}-\Delta\boldsymbol{K}\cdot\tilde{\boldsymbol{r}}_{1}-n\left(\theta+\tilde{\theta}_{1}\right)-\left(\theta+\tilde{\theta}_{R}\right)-\alpha\right)\tilde{F}\left(\tilde{R}\right)\\
\delta\rho_{BB} ^{(1)}\left(\boldsymbol{r}\right) & =-\frac{4}{\pi}\int d\tilde{\boldsymbol{r}}_{1}\Delta_{0}\left(\tilde{r}_{1}\right)\sin\left(\Delta\boldsymbol{K}\cdot\boldsymbol{r}-\Delta\boldsymbol{K}\cdot\tilde{\boldsymbol{r}}_{1}-n\left(\theta+\tilde{\theta}_{1}\right)+\left(\theta+\tilde{\theta}_{R}\right)-\alpha\right)\tilde{F}\left(\tilde{R}\right).\nonumber 
\end{align}

For a local radial function $\Delta_0\left(r\right)=-a\delta\left(r\right)$,  the integrals are numerically solved to provide  Fig.5 presented in the main text, where we have plotted $r^2 \left(\delta\rho_{AA} ^{(1)}+\delta\rho_{BB} ^{(1)} \right)$, without the oscillating radial function. 

\includegraphics[width=0.0001\textwidth]{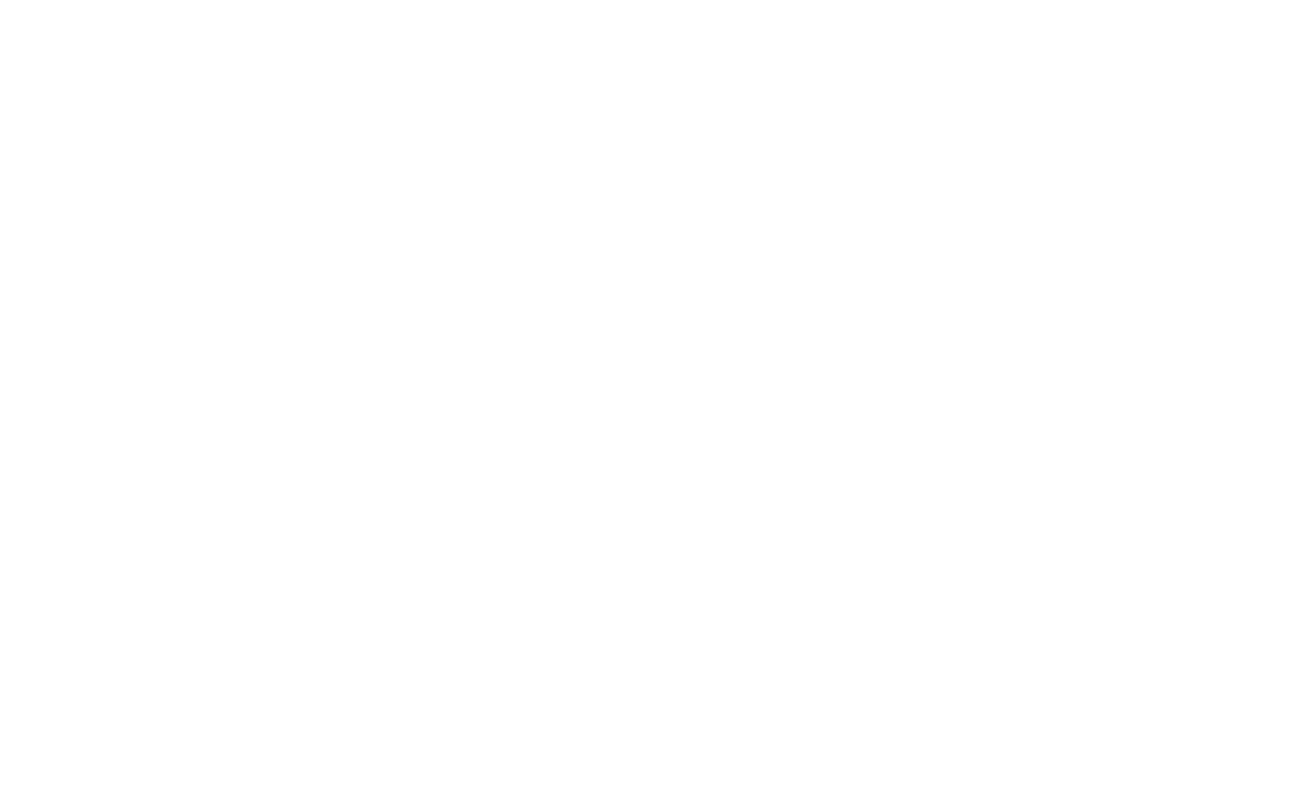}

%==================================================================================================== %  

%================ Figure ============================

%%%%%%% ======================================================================== %%%%%%%

\bibliographystyle{aipnum4-1}
% \bibliography{supplementary.bib}
